\documentclass[12pt]{article}
\usepackage[margin=3cm]{geometry}
\usepackage{xr-hyper}
\usepackage{amsmath,amssymb,amsfonts,amsthm}
\usepackage{setspace,algorithm,hyperref}
\usepackage[capitalize,nameinlink]{cleveref}
\usepackage{natbib}
\usepackage{graphicx}
\usepackage{comment,xcolor,soul}

\newtheorem{theorem}{Theorem}
\newtheorem{assumption}{Assumption}

\newtheorem{lem}{Lemma}
\newtheorem{remark}{Remark}

\newcommand{\ind}{\perp\!\!\!\!\perp}
\renewcommand{\d}{\mathrm{d}}
\newcommand{\V}{\mathbb{V}}
\renewcommand{\P}{\mathbb{P}}
\newcommand{\E}{\mathbb{E}}
\renewcommand{\mu}{Q}
\DeclareMathOperator*{\argmin}{argmin}

\def\ii{ I\{A_i = d^{\theta}_i \}}

\def\E{\mathbb{E}_{B_n}}

\renewcommand{\P}{\mathbb{P}}
\renewcommand{\d}{\mathrm{d}}
\newcommand{\score}{\mathcal{S}}

\newcommand{\prodK}{\prod_{t=1}^T \frac{I(A_t = d_t^{\theta})}{\pi_0^t(A_t = d_t^{\theta}|\bar H_t )}}
\newcommand{\Ij}{I(A_j = d_j^{\theta})}
\newcommand{\It}{I(A_t = d_t^{\theta})}
\newcommand{\pardevt}{\frac{\partial}{\partial \epsilon}}

\let\origtheassumption\theassumption

\title{Nonparametric assessment of regimen response curve estimators}
\date{}
\author{Cuong Pham$^{1}$, 
        Benjamin R. Baer$^{1,2}$, and
        Ashkan Ertefaie$^{1}$\\
        \vspace{6pt} 
        $^{1}$Department of Biostatistics and Computational Biology,\\
        University of Rochester Medical Center, Rochester, NY, USA \\
        $^{2}$School of Mathematics and Statistics, \\ University of St Andrews, St Andrews, Scotland}

\begin{document}

\maketitle
\doublespacing

\begin{abstract}

   Marginal structural models have been widely used in causal inference to estimate mean outcomes under either a static or a prespecified set of treatment decision rules. This approach requires imposing a working model for the mean outcome given a sequence of treatments and possibly baseline covariates. In this paper,  we introduce a {\it dynamic} marginal structural model that can be used to estimate an optimal decision rule within a class of parametric rules. Specifically, we will estimate the mean outcome as a function of the parameters in the class of decision rules, referred to as a {\it regimen-response curve}. In general, misspecification of the working model may lead to a biased estimate with questionable causal interpretability. To mitigate this issue, we will leverage risk to assess "goodness-of-fit" of the imposed working model. We consider the counterfactual risk as our target parameter and derive inverse probability weighting and canonical gradients to map it to the observed data. We provide asymptotic properties of the resulting risk estimators, considering both fixed and data-dependent target parameters. We will show that the inverse probability weighting estimator can be efficient and asymptotic linear when the weight functions are estimated using a sieve-based estimator. The proposed method is implemented on the LS1 study to estimate a regimen-response curve for patients with Parkinson's disease.

\end{abstract}

\section{Introduction}

Many diseases and disorders, including cancer, Parkinson's, substance abuse, and depression, require a sequence of treatments to address the evolving characteristics of the disease and the patient. Dynamic treatment regimes offer personalized treatment sequences to cater to these specific needs \citep{ robins1986new, robins1989analysis, robins1997causal,murphy2001marginal}. Such regimes consist of a predefined set of decision rules established before treatment initiation, which map the patient's time-varying characteristics to appropriate treatment options at each decision point. 

A key step toward estimating optimal treatment regimes is to estimate the mean potential outcome under a given regime. Marginal structural models have been widely used for this purpose, where a working model is imposed for the mean outcome given a sequence of treatments and possibly baseline covariates. However, the majority of the current literature focuses on finding an optimal regime among a prespecified set of regimes \citep{robins2000marginal, murphy2001marginal, shortreed2012estimating, ertefaie2016identifying}. Dynamic marginal structural models are alternatives that can be used to estimate optimal regimes within a class of parametric decision rules \citep{orellana2010dynamic, duque2021estimation}. This approach  estimates the mean outcome as a function of the parameters in the class of decision rules, referred to as a {\it regimen-response curve} \citep{ertefaie2023nonparametric}.  The quality of the estimated optimal regime and its causal interpretability depends on how closely the working model approximates the true regimen-response curve \citep{neugebauer2007nonparametric}. 



Cross-validation provides a powerful tool for model selection both in parametric and nonparametric settings. 
The asymptotic and finite sample properties of  cross-validated risk   have been extensively studied \citep{van2003unified,vaart2006oracle}. \cite{brookhart2006semiparametric}  illustrated how to utilize cross-validated risk for selecting the best nuisance parameters in a semi-parametric model. In this paper, we will use cross-validated risk to assess the performance of an imposed working model to estimate a regimen-response curve.  In this context, risk quantifies the proximity of the working model to the actual regimen-response curve. It is defined as the Euclidean distance between the working model and the potential mean outcome under all treatment regimes. The challenge lies in the fact that we do not  observe the potential mean outcome under a treatment regime. Hence, an inverse probability weighting (IPW) mapping is often utilized to estimate the risk of a  working model.

The consistency of the IPW estimators relies on a correctly specified propensity score model \citep{cole2008constructing, mortimer2005application, howe2011limitation}. This requirement is not always guaranteed when parametric methods are used. 
A solution to this problem is to use nonparametric methods. However, naively utilizing nonparametric methods to estimate the propensity score often leads to risk estimators that are not $\sqrt{n}$-consistent and suffer poor finite sample performance \citep{ertefaie2023nonparametric}. This is because the convergence rate of an IPW estimator relies on the convergence rate of the propensity score estimator, which is slower than the desired root-$n$ rate when nonparametric models are used. Additionally, inefficiency is a notable drawback of IPW estimators. In clinical trials that involve human subjects, efficiency is a critical consideration, as it is usually costly to have a large sample size.

Numerous approaches have emerged to address limitations in the IPW estimator. For instance, \cite{imai2015robust} proposed the covariate balancing propensity score, while \citet{yu2006double} introduced a double robust estimator. However, these methods were tailored for estimating the marginal structural model in a static treatment setting and were not applicable in identifying the optimal dynamic treatment setting.
We expand the literature by deriving the canonical gradient of the risk function and proposing a multiply robust estimator to estimate the risk of the dynamic marginal structural working model. This estimator maintains consistency even when the nuisance parameters were misspecified \citep{rotnitzky1998semiparametric, laan2003unified}. It achieves efficiency under the correct specification of all nuisance parameters. Furthermore, the multiply robust estimator allows for a slower convergence rate of the propensity score model, enabling the utilization of more flexible nonparametric methods for its modeling.
However, a drawback of the multiply robust estimator is its reliance on estimating numerous nuisance parameters. Employing parametric methods for their estimation can lead to inconsistency in some parameters, impacting the estimator's efficiency. Conversely, nonparametric methods ensure consistency but may become computationally burdensome, potentially yielding irregular estimators when some parameters are inconsistently estimated \citep{van2014targeted, benkeser2017doubly}.
Therefore, we also introduce a technique to refine the IPW estimator, enhancing its efficiency. Initially introduced by \cite{ertefaie2023nonparametric} for estimating causal treatment effects in point exposure settings, we extend this method to the longitudinal setting. 


The contribution of our proposed method beyond the existing literature can be summarized as follows. 
First, we derive the canonical gradient of the risk function corresponding to a dynamic marginal structural model in longitudinal settings. Our derivation leads to a multiply robust estimator of the risk function.  Second, we show that when the propensity score is estimated using a sieve-based approach, the resulting IPW estimator will solve the efficient influence function. This implies that the proposed IPW estimator achieves the nonparametric efficiency bound. Third, we show that, under certain rate conditions, the proposed multiply robust and sieve-based estimators are asymptotically linear for both fixed and data dependent risk functions.

\section{Problem Setting}

\subsection{Notation and setup}

Our data consists of $n$ i.i.d. samples of $O_{i} = \{ S_{1i},A_{1i}, \dots , S_{Ti},A_{Ti},Y_i \}^n_{i=1} \sim \P_0 \in \mathcal{M}$ where $\mathcal{M}$ is a nonparametric model. We denote $S_{t}$ as the vector of patient characteristics at time $t$ and $V \subseteq S_{1}$ as a subset of the baseline characteristics. The observed treatment is denoted $A_t \in \{0,1\}$ at time $t=1,\cdots,T$, and the outcome of interest is denoted $Y$. 
We use the overbar notation to denote the history of a variable. For example, $\bar S_t = (S_1, \dots, S_t)$ and  $\bar A_t = (A_1, \dots , A_t)$ are the covariates and past treatment up to time $t$. We let $\bar H_t = (\bar A_{t-1}, \bar S_t)$ be all the information up to time $t$. 
We denote $\P_n$ as the empirical distribution and define $\P f = \int f(o) \,\d \P(o)$ for a given function $f(o)$ and distribution $\P$.

A dynamic regime consists of a set of decision rules that assigns the treatment at each stage based on the information at and before that stage. We define the deterministic dynamic treatment rule $d^{\theta} = I( \theta_t^T S_{t}  > 0)$ to be the decision rule at time $t$. $\pi_0^t \equiv P(A_t = d_t^{\theta}| \bar H_t)$ is the true propensity score at time $t$ and $\pi_n^t$ denote the corresponding estimator. And, $Y^{\theta}$ denotes the potential outcome under the regime $\theta$. Suppose that we have the full data $X = \{(Y^\theta: d^\theta \in \mathcal{D}), V\} \sim P_X \in \mathcal{M}^F$  where $\mathcal{M}^F$ is a nonparametric full data model. 
We can then define the regimen response curve $m_0(\theta, V) := E_{P_X}(Y^{\theta}\mid V)$. 

\subsection{Cross-validated risk}


A dynamic marginal structural model can be used to approximate $m_0(\theta, V)$ by imposing a working model. We quantify the quality of the imposed working model using a cross-validated risk function. Let $B_n \in \{0,1\}^n$ denote the random partition of $\{1, \dots, n\}$ into two folds. A realization of $B_n=\{B_{n,1},B_{n,2},\cdots, B_{n,n}\}$ defines a particular split of the sample of $n$ observations
into a training set $\{i\in \{1,\cdots,n\}: B_{n,i}=0 \}$ and a validation set $\{i\in \{1,\cdots,n\}: B_{n,i}=1 \}$. Denote
by $\P_{n,b}^0$ and $\P_{n,b}^1$ the empirical distribution of the training sample and the validation sample. The type of cross-validation procedure is determined by the specific distribution of $B_n$. For example, in Monte Carlo cross-validation, where the sample is divided into training and validation sets with probability $p$, the distribution of $B_n$ puts mass $1/\binom{n}{np}$ on each binary realization of $B_n$. Another example is $B$-fold cross-validation, in which the distribution of $B_n$ puts mass $1/B$ on each possible realization. Throughout, we assume that this randomness is independent of all data. 

Consider an estimation rule $m_n(\theta, V) \equiv m_n(\theta, V \mid \P)$ that is fit using the distribution $\P$; we refer to $m_n$ as a marginal structural model (MSM) to be consistent with the literature, although it is the fit from a working model rather than a model itself. 
Let $m_{n,b}(\theta,V) \equiv m_n(\theta,V\mid \P_{n,b}^0)$ denote the MSM $m_{n}(\theta,V)$ evaluated at the empirical distribution $\P_{n,b}^0$, i.e. the training data.  
We denote $m^*(\theta, V)$ as the pointwise probability limit of $m_{n,b}(\theta,V)$. 
Note, $m^*$ does not necessarily equal the true regimen-response curve $m_0$. 

Define the full data conditional risk based on training observations as
\begin{align} \label{eq:trueriskmn}
    \Psi_{0}(m_n) 
    \equiv& \E \left[ \P_0 \left\{ \int \{Y^{\theta} - m_{n,b}(\theta,V) \}^2 \,\d F(\theta) \right\} \right], 
\end{align}
where $\E$ denotes an expectation over possible partitions and $F$ is a user-specified distribution function over the parameter $\theta$.  The averaging over the partitions in (\ref{eq:trueriskmn}) prevents the estimand from depending on any randomness in the selection of training data, thus making $\Psi_{0}(m_n)$ a data-adaptive estimand \citep{hubbard2016statistical}. 
 We define the full non-data-adaptive for a given regimen-response curve $m$ as $
    \Psi_{0, \text{nda}} (m)
    \equiv \P_0 \left\{ \int \{Y^{\theta} - m(\theta,V) \}^2 \,\d F(\theta) \right\}.$ 
The full data conditional risk, $\Psi_{0}(m_n)$ can be expressed in term of non-data-adaptive risk, $\Psi_{0, \text{nda}} (m)$, through the relation, $\Psi_{0}(m_n) = \E \left[ \Psi_{0, \text{nda}} (m_{n,b}) \right]$. In Section \ref{sec:cangra}, we develop efficiency theory for $\Psi_{0, \text{nda}} (m)$ of a given regimen-response curve, $m$. Then, we will use those results to construct a nonparametric asymptotically linear estimators for $\Psi_{0}(m_n)$.

\subsection{Inverse-probability weighted identification}

For a given marginal structural model $m_n$, the definition of the risk $\Psi_{0}(m_n)$ involves the potential outcome $Y^{\theta}$ which is not observed. However, under the following causal assumptions, it is identifiable using the observed data.
Recall that $\pi_0^t \equiv P(A_t = d_t^{\theta}| \bar H_t)$ is the true propensity score at time $t$.

\begin{assumption} (Causal assumptions) \label{assump:causal} 
\begin{enumerate}
    \item[]a. Consistency: $Y = Y^{\theta}$ if $\bar A_T = \bar d^{\theta}_T$ for all $\theta$; 
    \item[]b. Exchangeability: $A_t \ind Y^{\theta} \mid \bar H_t$ for all $\theta$ and all $t=1, \dots, T$;
    \item[]c. Positivity: $\prod_{t = 1}^T \pi_0^t(A_t = d^{\theta}_t \mid \bar H_t) > \epsilon$ almost surely, for all $\theta$ and some $\epsilon > 0$. 
\end{enumerate}
\end{assumption}

Assumption \ref{assump:causal}a states that the potential outcome under an observed regime equals the observed value. This assumption can be violated if the treatment assignment of a unit impacts the outcome of others \citep{hernan2020, kennedy2019nonparametric}. 
Assumption \ref{assump:causal}b states that there are no unmeasured confounders. This assumption holds in standard randomized studies. However, in observational studies it requires a rich enough set of characteristics $S_{1}, \dots, S_T$. 
Assumption \ref{assump:causal}c states that regardless of a subject's medical history, that person will have a nonzero probability of following any given treatment regime. 

\begin{lem}
\label{lem:identif}
    Let $m_n$ be an estimation rule for a regimen response curve. If Assumption \ref{assump:causal} holds, then the estimand is an observed-data functional, that is, 
    \begin{align*}
       \Psi_{0}(m_n) 
        &= \E \P_0 \left[ \left\{ \int \prod_{t = 1}^T \frac{I ( A_t = d_t^{\theta} ) }{\pi_{0}^t(A_t = d_t^{\theta} \mid \bar H_t )} \right\} \left\{Y - m_{n,b}(\theta, V) \right\}^2 \,\d F(\theta) \right].
    \end{align*}
\end{lem}
\subsection{Inverse-probability weighted estimator} \label{sec:ipwplug}

  Applying the plug-in approach to Lemma \ref{lem:identif}, we have the inverse probability weight (IPW) estimator for the full data conditional risk of a working model $m_n(\theta, V)$. 
       \begin{align*}
        \Psi_{n}^{\mathrm{IPW}}(m_n, \pi_n) 
        &= \E  \P^1_{n,b} \left[ \int \left\{ \prod_{t = 1}^T \frac{I(A_{t} = d_t^\theta)}{\pi^t_{n,b} (A_{t} = d_t^\theta \mid \bar A_{t-1}, \bar S_{t} )} \right\} \left\{Y - m_{n,b}(\theta, V) \right\}^2 \,\d F(\theta) \right],
    \end{align*}
where $\pi^t_{n,b}$ is the estimate of the propensity score applied to the training sample for the $b^{th}$ sample split.  
 The consistency and convergence rate of $\Psi_{n,m}^{\mathrm{IPW}}(\pi_n)$ depends entirely on the consistency and convergence rate of the estimators $\pi^t_{n}$ of $\pi^t_{0}$ for each time point $t=1,\dots,T$. It is common to estimate the propensity score using parametric models as it has the convergence rate of $O_p(n^{-1/2})$. 
 Alternatively, data-adaptive methods can be used to increase the flexibility of the model. However, due to a slow convergent rate of nonparametric methods, the IPW estimator could be biased. We can see that through the following decomposition. 

For a given marginal structural model $m(\theta, V)$, define the loss as
\begin{equation*}
    L_m( \pi) = \int \left\{ \prod_{t = 1}^T \frac{I(A_{t} = d_t^\theta)}{\pi^{t}} \right\} \left\{Y - m(\theta, V) \right\}^2 \,\d F(\theta). 
\end{equation*}
Then, we have 
\begin{align*}
    & \Psi_{n}^{\mathrm{IPW}}(m,\pi_n) - \Psi_{0}(m,\pi_0) 
    =  \P_n L_m(\pi_n) - \P_0 L_m(\pi_0) \\
    & \hspace{15mm} = (\P_n - \P_0) L_m(\pi_0) + \P_0\{ L_m(\pi_n) - L_m(\pi_0)\} + (\P_n - \P_0)\{L_m(\pi_n)  - L_m(\pi_0)\} \\
    & \hspace{15mm} = (I) + (II) + (III). 
\end{align*}

Assuming $\pi_n$ is a consistent estimator of $\pi_0$ and Donsker class condition, using standard empirical process theory, we can show that term (III) will converge to 0 at rate $n^{-1/2}$ \citep{van2000asymptotic}. Hence, $\Psi_{n,m}^{\mathrm{IPW}}$ is asymptotically linear only if term (II) also converges to 0 at the rate $n^{-1/2}$. Since that rate is not obtainable using nonparametric methods, term (II) will dominate the right hand side, causing $\Psi_{n,m}^{\mathrm{IPW}}$ to be biased. The same issue arises when $m_n(\theta, V)$ is a cross-fitted estimator of the underlying marginal structural model.

 \subsection{The highly adaptive lasso}

Our proposed approach for refining the inverse probability weighting estimator of $\Psi_0(m_n)$ utilizes the Highly Adaptive Lasso model (HAL). The detail of this process will be discussed in Section \ref{sec:uipw}. To provide the context, in this section, we will give a concise introduction to the HAL model and highlight its key characteristics. For an in-depth exploration of HAL, we refer our interested readers to \cite{benkeser2016highly, ertefaie2023nonparametric}
 
 The highly adaptive lasso is a nonparametric regression function that estimates infinite dimensional functions by constructing a linear combination of indicator functions that minimize a loss function  by constraining the $L_1$-norm complexity of the model. \citep{bibaut2019fast} shows that  highly adaptive lasso  estimators have a fast convergent rate of $n^{-1/3}$ (up to a log factor).     

Let $\mathbb{D}[0, \tau]$ be the Banach space of $d$-variate real-valued cadlag functions on a cube $[0, \tau] \in \mathbb{R}^d$. For a function $f \in \mathbb{D}[0, \tau]$, the sectional variation norm is defined as 
\begin{equation*}
    \|f\|_\nu=|f(0)|+\sum_{r \subset\{1, \ldots, d\}} \int_{0_r}^{\tau_r}\left|df_r(u)\right|,
\end{equation*}
where $f_r(u) = f\{u_1I(1 \in r), \dots, u_dI(d \in r) \}$ is the $r^{th}$ section of $f$ and the summation is over all subset of $\{1, \dots, d\}$. The term $ \int_{0_r}^{\tau_r}\left|df_r(u)\right|$ is the r-specific variation norm. 

For any $f \in \mathbb{D}[0, \tau]$ with $\|f\|_\nu<\infty$, \cite{gill1995inefficient} shows that $f$ can be represented as a linear combination of indicator functions. 
\begin{align}
f(x): & =f(0)+\sum_{r \subset\{1, \ldots, d\}} \int_{0_r}^{w_r} df_r(u) \nonumber \\
& =f(0)+\sum_{r \subset\{1, \ldots, d\}} \int_{0_r}^{\tau_r} I\left(u_r \leq x_r\right) df_r(u) .
\label{eq:halrep}
\end{align}

The latter integral in Equation \ref{eq:halrep} can be approximated using a discrete measure $F_n$ putting mass on the observed values of $X_r=\left(X_j: j \in r\right)$ denoted as $x_{r, i}$ for the $i$ the subject. We let $\beta_{r,i}$ denote the mass placed on each observation $x_{r,i}$ and $\phi_{r, i}=I\left(x_{i, r} \leq c_r\right)$. Then the approximated $f$ is
\begin{align*}
    \tilde{f}(x) &=f(0)+\sum_{r \subset\{1, \ldots, d\}} \sum_{i=1}^n I\left(x_{r, i} \leq c_r\right) f_{r, n}\left(d x_{i, r}\right)\\
    &=\beta_0+ \sum_{r \subset\{1, \ldots, d\}} \sum_{i=1}^n \beta_{r, i} \phi_{r, i}
\end{align*}

 In this formulation, $\left|\beta_0\right|+\sum_{r \subset\{1, \ldots, d\}} \sum_{i=1}^n\left|\beta_{r, i}\right|$ is an approximation of the sectional variation norm of $f$. Consequently, the minimum loss based highly adaptive lasso estimator $\beta_n$ is defined as
$$
\beta_n=\arg \min _{\beta:\left|\beta_0\right|+\sum_{r \subset\{1, \ldots, d\}} \sum_{i=1}^n\left|\beta_{r, i}\right|<\lambda} P_n L(\tilde{f}) .
$$
where $L(.)$ is a given loss function. This is a constrained minimization that corresponds to lasso linear regression with up to $n \times \left(2^d-1\right)$ parameters (i.e., $\left.\beta_{r, i}\right)$. Each value of $\lambda$ will give a unique HAL estimator. 

\section{Efficient multiply-robust estimation}

In this section, we develop a multiply robust (MR) estimator for the cross-validated risk $\Psi_{0}(m_n)$ of an estimation rule, or working model, $m_n$. 

\subsection{The Canonical Gradient of \texorpdfstring{$\Psi_{0, \text{nda}}(m)$}{Psi\_{0,\text{nda}}(m)}}
\label{sec:cangra}
   
To construct the MR estimator, we need to find the efficient influence function of $\Psi_{0, \text{nda}}(m)$. Influence functions are important because a regular and asymptotically linear estimator can be expressed as the empirical mean of the influence function plus some negligible term that converge to 0 at the rate of $n^{-1/2}$ or faster \citep{laan2003unified, tsiatis2006semiparametric}. The efficient influence function is the influence function whose variance corresponds to the efficiency bound. The influence function is unique in the nonparametric setting and equals the efficient influence function. 
The multiply robust estimator is defined using the efficient influence function, and we will show that it is asymptotically efficient when all the nuisance parameters converge to their true value at a rate of $n^{-1/4}$. This slower-than-parametric rate allows us to estimate the nuisance parameters with flexible nonparametric methods. 
Although the MR estimator is not efficient when the propensity score model is misspecified, it is still consistent when the outcome regression is correctly specified, or vice versa. 

The following result provides the nonparametric efficient influence function for the inner term in the definition of the cross-validated risk $\Psi_0(m_n)$. 
\begin{theorem}
\label{thr:inflfunc}
    Let $m$ be a given regimen-response curve. 
    If Lemma \ref{lem:identif} holds, the nonparametric efficient influence function for 
    \begin{equation*}
        \Psi_{0,\text{nda}}(m)
        \equiv \P_0 \int \{Y^{\theta} - m(\theta,V) \}^2 \,\d F(\theta)
    \end{equation*}
    is $\varphi(\pi_0, \mu_{0,m}; m) - \Psi_{0,\text{nda}}(m)$, where the uncentered term $\varphi(\pi_0, \mu_{0,m}; m)$ is 
    \begin{align*}
         \int \left[
            \left\{ \prod_{t  = 1}^T \frac{ I(A_t = d_t^{\theta})}{\pi^t_{0} } \right\} \{Y - m(\theta, V)\}^2
            - \sum_{t = 1}^T \frac{I(A_t = d^{\theta}_t) - \pi^t_{0} }{\pi^t_{0} }\mu^t_{0,m} \prod_{j = 1}^{t-1} \frac{I(A_t = d_t^\theta)}{\pi^j_{0}} 
        \right] \,\d F(\theta),
    \end{align*}
    and the nuisance parameter 
    \begin{equation*}
          \mu^t_{0,m} = \P_0 \left[ \left\{ \prod_{j = t + 1}^T \frac{ I(A_j = d_j^{\theta}) }{\pi^j_{0}} \right\} \{Y - m(\theta, V)\}^2 \middle| \bar S_t, \bar A_{t-1}, A_{t} = d^{\theta}_t  \right].
    \end{equation*}
\end{theorem}

The structure of the influence function in Theorem \ref{thr:inflfunc} involves a weighted average over $\theta$ of an inverse probability weighting term and an augmentation term for each of the time points. 
The function $\mu^t_{0,m}$ can be represented as a recursive sequential regression formula, which is useful in practice. 
 
\begin{remark} The function $\mu^t_{0,m}$ can be expressed recursively as 
\begin{equation*}
    \mu^t_{0,m} = \P_0 \left( \mu^{t+1}_{0,m} \mid \bar{S}_{t+1}, \bar{A}_{t-1}, A_t = d^{\theta}_t \right),
\end{equation*}
    for $t = 1, \dots,T$, where $\mu^{T+1}_{0,m} = Y$.
\end{remark}

\subsection{The estimator}

The form of the efficient influence function for $\Psi_{0, \text{nda}}$ may be used to construct an estimator for $\Psi_0(m_n)$ that is asymptotically efficient for certain choices of nuisance parameter estimators \citep{bickel1993efficient}. The estimator construction is more delicate than usual since the efficient influence function is for $\Psi_{0,\text{nda}}(m)$, with $m$ a given regimen-response curve, while the estimand is $\Psi_0(m_n) = \E \{ \Psi_{0, \text{nda}}(m_n(\cdot; \P^0_{n,b})) \}$. We first develop estimators as if $m$ were given, then we discuss how to adapt these estimators to instead target our estimand $\Psi_0(m_n)$. 

Consider estimation of $\Psi_{0,\text{nda}}(m)$ using the efficient influence function as a guide. Two common estimation approaches are to use one-step estimators, which add the the estimated efficient influence function to an initial estimator, and estimating-equation estimators, which treat the efficient influence function as an estimating function \citep{tsiatis2006semiparametric,kennedy2022semiparametric}. Since the efficient influence function in Theorem \ref{thr:inflfunc} only depends on $\Psi_{0,\text{nda}}(m)$ through its subtraction, each of these estimation strategies is equivalent. Thus we initially consider the estimator 
\begin{equation}
\label{eq:est1}
    \P_{n} \{ \varphi(\pi_{n}, \mu_{n, m}; m) \}, 
\end{equation}
where $\pi_n$ and $\mu_{n, m}$ are estimators for $\pi_0$ and $\mu_{0,m}$ respectively. Assuming these nuisance parameter estimators satisfy certain conditions, we can show that the above estimator is asymptotically efficient, i.e. has the efficient influence function of $\Psi_{0,\text{nda}}(m)$, found in Theorem \ref{thr:inflfunc}, as its influence function. 

Now consider estimation of our estimand $\Psi_0(m_n) = \E [ \Psi_{0, \text{nda}}\{ m_n(\cdot \mid \P^0_{n,b})\}]$; we construct our estimator by adapting the estimator in  (\ref{eq:est1}) for $\Psi_{0, \text{nda}}(m)$. A plugin-like approach readily produces the estimator 
\begin{equation}
\label{eq:est2}
    \E \{ \P_{n} \varphi(\pi_{n,b}, \mu_{n,b,m_{n,b}}; m_{n,b}) \},
\end{equation}
in which an outer average over folds is included, each fixed regimen-response curve $m$ is replaced with $m_{n,b}$ which is estimated in the training data $\P^0_{n,b}$ of the fold, and each nuisance parameter is replaced with estimators within the training data. 

The estimator in (\ref{eq:est2}) is almost a suitable estimator for the cross-validated risk $\Psi_0(m_n)$, however it still has one major deficiency which harms its performance. It averages over all observations, including those in the training set $\P_{n,b}^0$ as well as the validation set $\P_{n,b}^1$, which biases the estimator downwards. A corrected estimator, which we study throughout this paper, is  
\begin{align} 
\label{MRest}
    \Psi_{n}^{\mathrm{MR}}(m_n, \pi_n, \mu_{n,m_n}) 
    = \E \P^1_{n,b} \varphi(\pi_{n,b}, \mu_{n,b, m_{n,b}}; m_{n,b}),
\end{align}
where $\pi_{n,b}$ and $\mu_{n,b, m_{n,b}}$ are estimates of the  nuisance parameters $\pi_0$ and $\mu_{n,m_0}$ calculated with the training sample for the $b^\text{th}$ sample split.


The next result establishes that $\Psi_{n}^{\mathrm{MR}}$ is an asymptotic efficient estimator for the estimand $\Psi_0(m_n)$. Before stating it, we make two assumptions. 

\begin{assumption} \label{assump:vrate2}
    For all $t = 1,\dots,T$, the nuisance parameters $\pi^t_{n}$ and $\mu^t_{n,m_{n,b}}$ satisfy  $\E\int  \|\pi^t_{0} - \pi^t_{n,b} \|_{L_2(\P_0)} \|\mu^t_{0,m_{n,b}} - \mu^t_{n,m_{n,b}} \|_{L_2(\P_0)} dF(\theta) = o_p(n^{-1/2})$.
\end{assumption}

   
\let\theassumption\origtheassumption

\begin{assumption}\label{assump:mn-rate}
   The estimator $m_{n,b}$ satisfies $\int \|m_{n,b}(\theta; V) - m^*(\theta; V) \|_{L_2(\P_0)} dF(\theta) = o_p(n^{-1/4})$. 
\end{assumption}

Assumption \ref{assump:vrate2} states that the estimators $\pi_n^t$ and $\mu^t_{n,m_{n,b}}$ need to converge to its true value at a rate of $o_p(n^{-1/4})$ for any $\theta$. This assumption holds when these nuisance parameters are estimated using the highly adaptive lasso. Moreover, if either $\pi_0^t$ or $\mu^t_{n,m_{n,b}}$ converge to its true value at the rate of $O_p(n^{-1/2})$, we only require consistency for the other nuisance parameter. The rate of $O_p(n^{-1/2})$ can be achieved using parametric methods. Assumption \ref{assump:mn-rate} is also a relatively mild assumption. It only requires the fitted marginal structure model, $m_{n,b}$, to have a limit $m^*$. And $m_{n,b}$ converges to $m^*$ at the rate of $o_p(n^{-1/4})$, enabling the utilization of both parametric and nonparametric methods to estimate the marginal structure model.    
   
\begin{theorem}  
\label{thr:flexiblem}
 If Assumptions \ref{assump:causal} holds, then the estimator $\Psi_{ n}^{\mathrm{MR}}(m_n, \pi_n, \mu_{n,m_n})$ satisfies 
   \begin{equation*}
       \Psi_{ n}^{\mathrm{MR}}(m_n, \pi_n, \mu_{n,m_n}) - \Psi_{0}(m_n) = (\P_n - \P_0)\varphi(\pi_0, \mu_{0,m^*}; m^*) + R_n,
   \end{equation*}
   where $R_n$ is defined in Section \ref{proof:thrm2} of the Supplement Material. 
   If Assumption \ref{assump:vrate2} additionally holds, then $R_n = o_p(n^{-1/2})$ and 
   \begin{equation*}
       \sqrt{n} \left\{ \Psi_{n}^{\mathrm{MR}}(m_n,\pi_n, \mu_{n,m_n}) - \Psi_{0}(m_n) \right\} 
       \rightsquigarrow \mathcal{N} \left[ 0, \V_0\{\varphi(\pi_0, \mu_{0,m^*}; m^*)\} \right].
   \end{equation*}
\end{theorem}

Theorem \ref{thr:flexiblem} shows that as the fitted working marginal structural model $m_{n,b}^0$ converges to its limiting value $m^*$ at the rate of $o_p(n^{-1/4})$, the multiply robust estimator is asymptotic linear with respect to the limit of $m_n$. The centering term in Theorem \ref{thr:flexiblem} is the data-adaptive risk $\Psi_{0}(m_n)$ that depends on the estimator $m_n$, the true propensity score and outcome regression. 

When two working models are under considerations, the results of Theorem \ref{thr:flexiblem} allow us to compare their estimators, $m_{n,1}$ and $m_{n,2}$, by applying the asymptotic linearity result once for each estimator. We may ascertain the quality of the estimators by comparing their risks $\Psi_{0}(m_{n,1})$ and $\Psi_{0}(m_{n,2})$ by testing the informal null hypothesis that these risks are equal. 

In Theorem \ref{thr:flexiblem}, we study the difference $\Psi^{\mathrm{MR}}_{n}(m_n, \pi_n, \mu_{n,m_n}) - \Psi_{0}(m_n)$. This is because $\Psi^{\mathrm{MR}}_{n}(m_n, \pi_n, \mu_{n,m_n}) - \Psi_{0}(m^*)$ is asymptotically linear under much more restrictive assumptions. In the following result, we study the asymptotics of $\Psi^{MR}_n(m_n,\pi_n, \mu_{n,m_n}) - \Psi_{0}(m^*)$ to replace the data-adaptive target parameter $\Psi_{0}(m_n)$ with the fixed target parameter $\Psi_{0}(m^*)$ in the conclusions of Theorem \ref{thr:flexiblem}. 
\begin{theorem}
\label{cor:psi-m-star}
   If Assumptions \ref{assump:causal} and \ref{assump:mn-rate} hold and additionally $m^*$, the limit of $m_n$, equals the true minimizer of $\Psi_{0}(m)$ over $m \in \mathcal{M}$ (i.e., $m^*=m_0$), then 
   \begin{align*}
       \Psi^{MR}_{n}(m_n,\pi_n, \mu_{n,m_n}) - \Psi_{0}(m^*) 
       = (\P_n - \P_0)\varphi(\pi_0, \mu_{0,m_0}; m_0) + R_n,
   \end{align*}
   for $R_n$ defined in Section \ref{proof:thrm3} of the Supplementary Materials. If additionally Assumption \ref{assump:vrate2} holds, then $R_n = o_p(n^{-1/2})$ and 
   \begin{equation*}
       \sqrt{n} \left\{ \Psi_{n}^{\mathrm{MR}}(m_n, \pi_n, \mu_{n,m_n}) - \Psi_{0}(m^*) \right\}  
       \rightsquigarrow \mathcal{N} [ 0, \V_0\{\varphi_{m^*}(\pi_0, \mu_{0,m^*})\} ].
   \end{equation*}
\end{theorem}
The result of Theorem \ref{cor:psi-m-star} can be used to compare two different modeling approaches known to be consistent with $\|m_{n,1}-m_0\|_2=o_p(n^{-1/4})$ and $\|m_{n,2}-m_0\|_2=o_p(n^{-1/4})$. Despite having the same asymptotic variance, the approximated variance based on $\varphi(\pi_0, \mu_{0,m_{n,1}}; m_{n,1})$ and $\varphi(\pi_0, \mu_{0,m_{n,2}}; m_{n,2})$ may differ. As an illustration, one could consider comparing the super learner with a library that includes HAL alongside other parametric and nonparametric methods, against a HAL-only model.
\section{Efficient plugin estimation}

\subsection{The estimator}
\label{sec:uipw}

Although doubly robust procedures facilitate the use of data
adaptive techniques for modeling nuisance parameters, the resultant estimator
can be irregular with large bias and a slow rate of convergence when either of
the nuisance parameters is inconsistently estimated \citep{van2014targeted,benkeser2017doubly}. To mitigate this issue, we propose a sieve based IPW estimator called UIPW where the propensity scores are estimated using an undersmoothed highly adaptive lasso. Define the estimator as
\begin{align}
    \Psi_{ n}^{\mathrm{UIPW}}(m_n,\pi_{n,\lambda}) = \E \P^1_{n,b} \left[ \int \left\{ \prod_{t = 1}^T \frac{I(A_{t} = d_t^\theta)}{\pi^t_{n,b,\lambda} (A_{t} = d_t^\theta \mid \bar H_t )} \right\} \left\{Y - m_{n,b}(\theta, V) \right\}^2 \,\d F(\theta) \right],
\end{align}
where $\pi^t_{n,b,\lambda}$ is the highly adaptive lasso estimate of $\pi^t_{0}$ obtained using the b$^{th}$ sample split for a given tuning parameter $\lambda$. Smaller values of $\lambda$  imply a weaker penalty, thereby undersmoothing the fit.

Our theoretical results show that for specific choices  $\lambda$, the function $\Psi_{m_n, n}^{\mathrm{UIPW}}(\pi_{n,\lambda})$ can solve the efficient influence function presented in Theorem \ref{thr:inflfunc}. This implication indicates that the proposed undersmoothed highly adaptive lasso estimator achieves nonparametric efficiency. This accomplishment relies on the satisfaction of two key assumptions:
 
\begin{assumption}[Regularity of Functions]
    \label{assum:cadlag}
    $\mu^t_{0,m^*}$ and $\pi^t_0$ for any $1 \le t \le T$ are cadlag functions with finite sectional variation norm.
\end{assumption}

\begin{assumption}[Approximation of Functions]
    \label{assum:quarterneighbor}
     For $1 \le t \le T$, let $f_t = \frac{\mu^t_{0,m^*}} {\pi^t_{0}}\prod_{j=1}^{t-1} \frac{I(A_t = d^{\theta}_t)}{\pi^j}$, and $\tilde f_t$ be the projection of $f_t$ onto the linear span of the basic functions $\phi^t_{r,i}$ in $L^2(P)$. Here, $\phi^t_{r,i}$ represents the basis functions generated by the highly adaptive lasso fit of $\pi^t_{n,b,\lambda}$. We assume that $\|f_t - \tilde f_t \|_{L_2(\P_0)} = O_p(n^{-1/4})$. 
\end{assumption}

 Assumption \ref{assum:cadlag} typically holds in practical applications because the set of cadlag functions with finite sectional variation norms is extensive.Assumption \ref{assum:quarterneighbor}, on the other hand, posits that the generated basis functions in $\pi^t_{b, \lambda}$ are sufficient to approximate $f_t$ within an $O_p(n^{-1/4})$ neighborhood of $f_t$. Without undersmoothing, the basis functions can effectively approximate $\pi^t_{b, \lambda}$ when $\pi^t_0$ is as complex as $f_t$. However, when the complexity of $f_t$ surpasses that of $\pi^t_{b, \lambda}$, more undersmoothing is necessary. 

\begin{theorem}[Efficiency of UIPW estimator] Let
\label{thr:uipw}
\begin{align*}
     D^t_{CAR}(\pi,\mu) = \int \left\{ \frac{I(A_t = d^{\theta}_t) - \pi^t }{\pi^t }\mu^t_{m} \prod_{j = 1}^{t-1} \frac{I(A_t = d_t^\theta)}{\pi^j} \right\} dF(\theta). \label{eq:dcar}
\end{align*}

    Under Assumptions \ref{assump:causal},\ref{assum:cadlag},\ref{assum:quarterneighbor} and $\E\int \|\pi_{n,b}-\pi_0\|_2. \left(\|\pi_{n,b}-\pi_0\|_2  + \|m_{n,b}-m^*\|_2 \right) dF(\theta) =o_p(n^{-1/2})$, there exists a set of $\lambda_{opt} = \{\lambda^1_{opt}, ..., \lambda^T_{opt}\}$ such that 
    $\E\P_n \sum_{t = 1}^T D^t_{CAR}(\pi_{b,\lambda_{opt}},\mu_{0, m^*})=o_p(n^{-1/2})$
    and the estimator $\Psi^{UIPW}_n({m_n}, \pi_{n,\lambda_{opt}})$ is asymptotically efficient and satisfies
 \begin{equation*}
           \Psi^{UIPW}_{n}(m_n,\pi_{n,\lambda_{opt}}) - \Psi_{0}(m_n,\pi_0) = (\P_n - \P_0)\varphi(\pi_0, \mu_{m^*, 0}; m^*) + o_p(n^{-1/2}). 
 \end{equation*}
    
\end{theorem}

\subsection{Undersmoothing criteria in practice}

As discussed in Section \ref{sec:ipwplug}, without undersmoothing, the estimator $\Psi_{m_n, n}^{\mathrm{UIPW}}$ may fail to be asymptotically linear. An IPW estimator will be efficient if it solves the efficient influence function. Theorem \ref{thr:uipw} shows that the latter is achieved when $\P_n \sum_{t = 1}^T D^t_{CAR}(\pi_{n,\lambda},\mu_{0, m^*})=o_p(n^{-1/2})$. The theoretical rate might not provide practical guidance due to the unknown constant. However, given the finite sample at hand, one can achieve the best performance by selecting a tuning parameter that minimizes the  $\P_n \sum_{t = 1}^T D^t_{CAR}(\pi_{n,\lambda},\mu_{0, m^*})$. Specifically, we define
\begin{align}
\label{eq:dcar}
    \lambda_{n,t} = \argmin_{\lambda} \left| B^{-1} \sum_{b=1}^B \P^1_{n,b} D^t_{CAR}(\pi_{n,b,\lambda},\mu_{n,b,m_b}) \right|, 
\end{align}
where $\mu_{b,m_b}$ is a cross-validated highly adaptive lasso estimate of $\mu_{0, m^*}$ with the $L_1$-norm bound based on the global cross-validation selector.

The criterion (\ref{eq:dcar}) relies on knowledge about the efficient influence function. However, in some cases, deriving this function can be intricate (e.g., longitudinal settings with many decision points) or may not have a closed-form solution (e.g., bivariate survival probability in bivariate right-censored data) \citep{van1996efficient}. To this end, we propose an alternative criterion, drawing from the approach in \cite{ertefaie2023nonparametric}, that does not require the efficient influence function. Define
\begin{equation}\label{eq:score}
 \tilde \lambda_{n,t} = \argmin_{\lambda}  B^{-1} \sum_{b=1}^B \left[ \sum_{(r,i) \in
 \mathcal{J}_n} \frac{1}{ \lVert \beta_{t,n,\lambda,b} \rVert_{L_1}}
 \bigg\lvert \P_{n,b}^1 \tilde \Omega_{r,i}(\phi,\pi^t_{n,b,\lambda})
 \bigg\rvert \right],
\end{equation}
in which $\lVert \beta_{t,n,\lambda,b} \rVert_{L_1} = \lvert \beta_{t,n,\lambda,0}
\rvert + \sum_{r \subset\{1, \ldots, d\}} \sum_{j=1}^{n} \lvert
\beta_{t,n,\lambda,r,i} \rvert$ is the $L_1$-norm of the coefficients
$\beta_{t,n,\lambda,r,i}$ in the highly adaptive estimator $\pi^t_{n,\lambda}$
for a given $\lambda$, and $\tilde \Omega_{r,i}(\phi, \pi^t_{n,\lambda,v}) =
\phi_{r,i}(\bar H_{t}) \{A_{j} - \pi^t_{n,\lambda,v}(1 \mid \bar H_{t})\}\{\pi^t_{n,\lambda,v}
(1 \mid \bar H_{t}))\}^{-1}$. 

The main drawback of undersmoothing is the decreased convergence rate of the highly adaptive lasso fit. Our asymptotic linearity proof of $\Psi_{m_n, n}^{\mathrm{UIPW}}(\pi_{n,\lambda})$ requires $\int \| \pi^t_{n} -\pi^t_{0}\|_2 dF(\theta) = o_p(n^{-1/4})$ which is slower than the cross-validated based highly adaptive lasso fit of $O_p(n^{-1/3})$. This ensures that we can tolerate a certain degree of undersmoothing without compromising our desired asymptotic properties. Let $J$ be a number of features included in fit (i.e., $J=|\mathcal{J}_n|$). \cite{van2023higher} showed that the convergence rate of a highly adaptive lasso is $(J/n)^{1/2}$. Hence to ensure the required $o_p(n^{-1/4})$, we must choose a $J$ such that $J<n^{1/2}$ and let $\lambda_{n,J}$ be the corresponding $L_1$ norm. We then propose our score based criterion as 
\begin{align} \label{UIPWScore}
    \lambda'_{n,t} = \max(\lambda_{n,J}, \tilde \lambda_{n,t}).
\end{align}
In practice, the score-based criterion tends to undersmooth too much (i.e., pick a very small $\tilde \lambda_{n,t}$), and the max operator helps us to mitigate this issue. 

\section{Simulation Studies}

\subsection{Setting}
In our simulation study, we evaluate and compare the performance exhibited by our proposed estimators, MR and UIPW to the conventional IPW estimator. We consider a two-stage treatment in our simulation (T = 2). The baseline covariates $S_{1,t=1}, S_{2,t=1}$ are generated from a normal distribution with the mean equals 0 and the standard deviation equals 0.5. The time-dependent covariate $S_{1,t}, S_{2,t}$ are function of the preceding $S_{1, t-1}, S_{2,t-1}$. In particular, 
 \begin{align*}
            &S_{1,t} = \text{Max}[\text{Min}\{0.9(2A_{t - 1})S_{1,t-1} + 0.05S_{1,t-1}S_{2,t-1} + N(0, 0.5),8\},-4] \\
            &S_{2,t} = \text{Max}[\text{Min}\{0.9(1-2A_{t - 1})S_{1,t-1} + 0.05S_{1,t-1}S_{2,t-1} + N(0, 0.5),8\},-4]
 \end{align*}
 The outcome $Y =  S_{1,T} - S_{2,T} + A_{T}S_{1,T} + N(0,0.1^2)$ is a function of  $S_1, S_2$ in the last stage. The decision rule is given as $d^{\theta}_t = I\{S_{1,t} < \theta\}$ which only depends on $S_{1,t}$. We generate $\theta$ from a normal distribution $f(\theta) = N(0, 0.1^2)$. The treatment $A_t$ is generated from a Bernoulli distribution with $P(A_t = 1|S_{1,t}, S_{2,t}) = \text{expit}(0.1 + 0.2S_{1,t})$. Additional simulations where $A_t$ is randomized can be found in Section \ref{sec:addsim} of the Supplementary Material.  

 We consider two variants of the UIPW estimator, differing in how the tuning parameter $\lambda$ is selected. UIPW-Dcar chooses $\lambda$ based on the criterion \ref{eq:dcar}, while UIPW-Score chooses $\lambda$ based on the criterion \ref{UIPWScore}. The nuisance parameters, $\mu^t_{n,m_n}$ and $\pi^t_{n}$, is estimated using the highly adaptive lasso. We impose the parametric working model $m = E[Y|\theta] = \beta_0 + \beta_1\theta$. We perform 360 simulations on the sample sizes 250, 500, 850, and 1000. In Section \ref{sec:addsim} of the Supplementary Material, we present additional simulations when the working model $m$ is estimated using other regression models, both parametric and nonparametric.
\subsection{Efficiency of the Proposed Estimators} \label{sim:5.2}

\begin{figure}[ht]
    \centering
    \includegraphics[width = \textwidth]{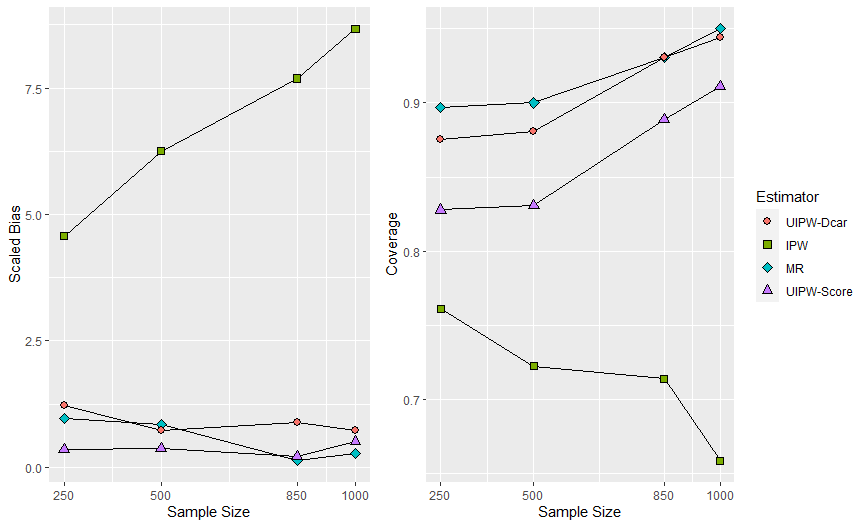}
    \caption{Bias and Coverage plots}
    \label{fig:plot1}
\end{figure}

 We showcase the effectiveness of our proposed estimators via coverage plots, as seen in the right panel of Figure \ref{fig:plot1}. Notably, the coverage of IPW consistently falls below 77\% at a sample size of 250 and diminishes further with larger sample sizes. Conversely, the MR and UIPW-Dcar estimators maintain stable coverage around 90\% at small sample sizes and achieving 93\% and 95\% at sample sizes of 850 and 1000, respectively.
Regarding UIPW-Score, its initial coverage hovers around 85\% at sample sizes of 250 and 500, but rapidly climbing to approximately 92\% at a sample size of 1000. 

The left panel of Figure \ref{fig:plot1} provides insight into the scaled bias, calculated by multiplying the bias by the square root of the sample size.  The MR estimator displays the smallest scaled bias. Additionally, as the sample size increases, our proposed estimators exhibit a declining trend in scaled bias, indicating convergence towards the true value at a rate of $o_p(n^{-1/2})$.

Conversely, the scaled bias of the IPW estimator shows an upward trend with increasing sample size, signifying a slower convergence rate than $o_p(n^{-1/2})$. This slower convergence is attributed to the bias term in the IPW estimator, which does not diminish rapidly enough when estimating the propensity score model using a nonparametric method.

\section{Data Application}

We implement our proposed methods to determine the optimal treatment strategy for individuals with Parkinson's Disease, focusing on the LS1 study. The LS1 study enrolled patients diagnosed with Parkinson's within five years, being treated with either Levodopa or DRA drug class \citep{kieburtz2015effect}. However, given the prevalent use of Levodopa as the first line therapy, we concentrate solely on patients already on Levodopa before the study. The study includes follow-ups at three months initially and annually thereafter, with our primary outcome ($Y$) being the UPDRS-III score two years after the start of the LS1 study. UPDRS-III is a measure of motor function of a patient, ranging from 0 to 100.

Within our sample of 238 patients, the UPDRS-III score ranges from 3.00 to 68.00, with mean and median values around 20.00, and the first and third quantiles are at 13.00 and 26.00, repspectively. We focus on two decision points for treatment adjustment: at three months and a year into the study (i.e., $T=2$). Treatment decisions involve high (total Levodopa $>$ 300 mg/day) or low ( $\le$ 300 mg/day) doses (i.e., $A_t \in \{high, low\}$). Our decision rule, $d^{\theta}$, assigns high or low doses based on whether the UPDRS-III score exceeds the threshold $\theta$, i.e $d^{\theta} = I(\theta > \text{UPDRS-III})$.

We seek to find $\theta_{opt} = \argmin_{\theta \in [0,30]} E[Y^{\theta}]$ where the choice of $\theta$'s range [0,30] aligns with our data support. We consider three marginal structural models:

Model 1: Quadratic linear regression in $\theta$; i.e $E[Y\mid \theta] = \beta_0 + \beta_1\theta + \beta_2\theta^2$. 

Model 2: Highly adaptive lasso model with zero order splines

Model 3: Highly adaptive lasso model with smoothing.

We will employ cross-validation to prevent overfitting. For risk calculation, we employ four estimators: DR, IPW, and UIPWs (Dcar and Score). These models leverage the highly adaptive lasso to fit both the propensity score model $\pi^t$ and the nuisance parameter $Q^t$, using multiple covariates.

\begin{figure}
    \centering
    \includegraphics[width=0.7\linewidth]{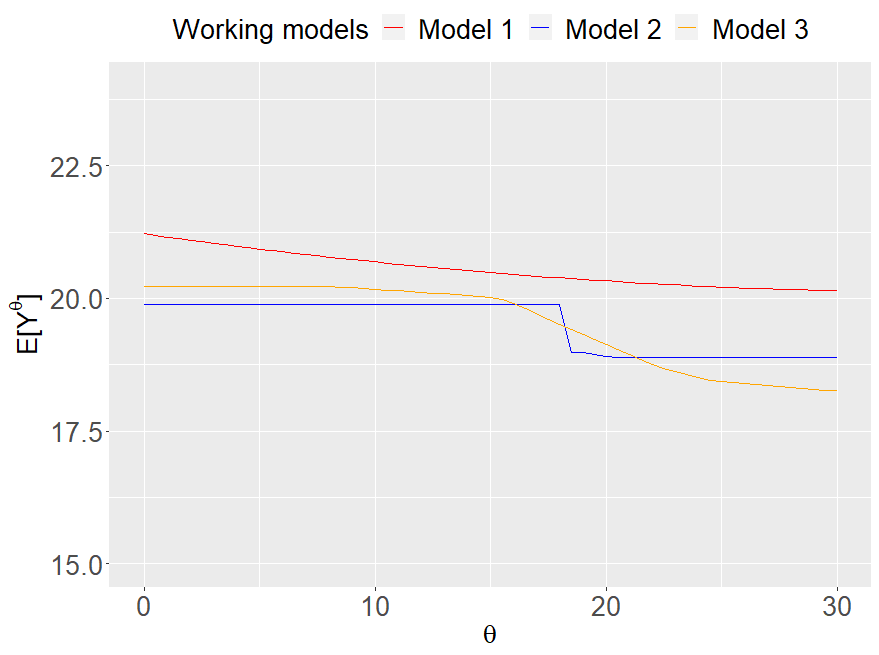}
    \caption{Working models estimated using 5-fold cross validation}
    \label{fig:msm-plots}
\end{figure}

The risk estimates in Table \ref{dataapp:risk} indicate that the linear regression model (Model 1) consistently yields the highest risk among all estimators. However, the estimated risks of the highly adaptive lasso models (Model 2 and 3) are close to the risks of the linear regresion model. Examining Table \ref{dataapp:opttheta}, optimal $\theta$ values across folds mostly converge around 30, suggesting minimal improvement with higher Levodopa doses. Figure \ref{fig:msm-plots} presents marginal structural model plots, indicating minimal marginal benefits or none with increased Levodopa doses.  

Dyskinesia and motor fluctuation are two major side effects of Levodopa treatment. To examine the association between Levodopa doses (i.e., high vs low) with these side effects, we fit a logistic model that includes the binary dosing of Levodopa as a dependent variable and the presence of dyskinesia or motor fluctuation recorded in UPDRS-IV after a year of treatment initiation as an independent variable. Our analysis indicates that the odd ratio of developing adverse events in the high dose is 1.6 times that in the low dose. This finding is consistent with findings from other studies that a high dose of Levodopa is associated with dyskinesia and motor fluctuation \citep{thanvi2007levodopa}. Hence, our analysis suggests favoring low-dose Levodopa for early-stage Parkinson's patients. It is because high dose of Levodopa has minimal improvement in UPDRS-III scores but increases likelihood of having adverse events. 

\begin{table}[ht]
\centering
\caption{Estimated risk with different risk estimators}
\begin{tabular}{crrrr}
 \hline
 Model  & DR  &  UIPW-Dcar & IPW  & UIPW-Score \\
\hline
Model 1  & 114.19 & 199.33 & 199.33 & 199.33 \\
Model 2  & 125.64 & 204.16 & 204.16 & 204.16 \\
Model 3  & 121.94 & 203.02 & 203.02 & 203.02\\

   \hline
\end{tabular}

\label{dataapp:risk}
\end{table}

\begin{table}[ht]
\centering
\caption{Optimal points with each fold}
\begin{tabular}{rrrrrr}
  \hline
Fold &  Model 1 & Model 2 & Model 3  \\ 
  \hline
Fold 1 &  19.5 & 0.0 & 30.0  \\
Fold 2 & 30.0 & 18.5 & 30.0  \\
Fold 3 & 30.0 & 18.5 & 22.5  \\
Fold 4 & 17.0 & 21.5 &  25.0  \\
Fold 5 & 30.0 & 18.5 & 30.0    \\
\hline
\end{tabular}

\label{dataapp:opttheta}
\end{table}

\section{Discussion}

This paper introduces two efficient estimators for assessing the risk associated with a marginal structural model. The first is a multiply robust estimator, validated to achieve efficiency when all nuisance parameters are correctly specified. It remains consistent even when the propensity score or the mean outcome model is accurately specified. The second is a sieve IPW estimation leveraging Highly Adaptive Models to attain efficiency.

Our work has several promising directions for extension. While our method is adaptable to studies with multiple stages, the assumption of positivity often fails in scenarios with a large number of stages. A potential modification involves transitioning from a deterministic to a stochastic decision rule where a probability is assigned to each possible treatment such that higher probabilities reflect the higher chance of being the optimal treatment. Stochastic  rules are of interest because their carry more information about the relative optimality of a treatment compared to the other treatment choices.

To facilitate the comparison of  several marginal structural models, one can use our asymptotic linearity results to identify the set of best models that contains the true best model with a given probability. This will necessitate the generalization of the multiple comparison with the best approach to our specific setting.  
Dynamic marginal structural models can be used with time-to-event data by considering structural Cox or accelerated failure time models for the potential outcomes. 

\section*{Acknowledgements}
CP, BRB, and AE were supported by the National Institute of Neurological Disorders and Stroke (R61/R33 NS120240).
\vspace*{-8pt}

\bibliographystyle{plainnat}
\bibliography{Risk-compare}

\begin{center}
{\large\bf SUPPLEMENTARY MATERIAL}
\end{center}

\section{Proof of Lemma 1}

The lemma follows in the usual way. For any given regimen-response curve $m$, we see that
\begin{align*}
    & \P_0 \left[\int \prod_{t = 1}^T \frac{I ( A_t = d_t^{\theta} ) }{\pi^t_{0}(A_t = d_t^{\theta} \mid \bar H_t )} \left\{Y - m(\theta, V) \right\}^2 \,\d F(\theta) \right] \\
    & \hspace{15mm} = \P_0 \left[\int \prod_{t = 1}^T \frac{I ( A_t = d_t^{\theta} ) }{\pi^t_{0}(A_t = d_t^{\theta} \mid \bar H_t )} \left\{Y_{\theta} - m(\theta, V) \right\}^2 \,\d F(\theta) \right] \\
    & \hspace{15mm} = \P_0 \left[\int \frac{\pi^T_{0}(A_t = d_t^{\theta} \mid \bar H_t, Y_{\theta})}{\pi^T_{0}(A_t = d_t^{\theta} \mid \bar H_t )} \prod_{t = 1}^{T-1} \frac{I ( A_t = d_t^{\theta} ) }{\pi^t_{0}(A_t = d_t^{\theta} \mid \bar H_t )} \left\{Y_{\theta} - m(\theta, V) \right\}^2 \,\d F(\theta) \right] \\
    & \hspace{15mm} = \P_0 \left[\int \prod_{t = 1}^{T-1} \frac{I ( A_t = d_t^{\theta} ) }{\pi^t_{0}(A_t = d_t^{\theta} \mid \bar H_t )} \left\{Y_{\theta} - m(\theta, V) \right\}^2 \,\d F(\theta) \right] \\
    & \hspace{15mm} = \dots \\
    & \hspace{15mm} = \P_0 \left[\int \left\{Y_{\theta} - m(\theta, V) \right\}^2 \,\d F(\theta) \right],
\end{align*}
where the second equality applies iterated expectation and the dots formally follow through induction. 

Now, recall that 
\begin{equation*}
    \Psi_{0}(m_n) 
    = \mathbb{E}_{B_n} \P_0 \int \{Y^{\theta} - m_{n,b}(\theta,V) \}^2 \,\d F(\theta)
\end{equation*}
is an average over folds. Taking the regimen-response curve $m$ to be the curve $m_{n,b}$ fitted to the training data $\P^1_{n,b}$ and $\P_0$ to be the expectation with respect to a (new) test point, our previous calculation shows that
\begin{align*}
    \P_0 \left[\int \prod_{t = 1}^T \frac{I ( A_t = d_t^{\theta} ) }{\pi^t_{0}(A_t = d_t^{\theta} \mid \bar H_t )} \left\{Y - m_{n,b}(\theta, V) \right\}^2 \,\d F(\theta) \right] 
    = \P_0 \left[\int \left\{Y_{\theta} - m_{n,b}(\theta, V) \right\}^2 \,\d F(\theta) \right].
\end{align*}
The result now follows by averaging over folds.

\section{Proof of Theorem 1}

  The target parameter $\Psi_{0,\text{nda}}(m)$ is a pathwise differentiable function. Hence, the canonical gradient of $\Psi_{0, \text{nda}}(m)$ in a nonparametric model $\mathcal{M}$ corresponds to the efficient influence function of a regular, asymptotically linear (RAL) estimator. Let $\Psi_{\epsilon, \text{nda}}(m) = P_{\epsilon} L_m(\pi)$ where $P_{\epsilon}=(1+\epsilon \mathcal{S})P_0$ is a perturbed data distribution.   The canonical gradient $\varphi$ is the one that satisfies the following equation: 
  \[
\frac{\partial}{\partial \epsilon} \Psi_{\epsilon, \text{nda}}(m) \mid_{\epsilon=0} = E\{\varphi(m) \mathcal{S}(O;\epsilon)\}\mid_{\epsilon=0},
  \]
  where $\score(O;\epsilon) = \frac{\partial}{\partial \epsilon} \log f(O;\epsilon)$.

 Let $\pi_\epsilon$ denote the perturbed propensity score. Then, 
\begin{align}
    \frac{\partial}{\partial \epsilon} \Psi_{\epsilon ,\text{nda}}(m) \mid_{\epsilon=0} &= \mathbb{E}\int \Big\{\prodK\{Y - m(\theta,V)\}^2 \score(\mathcal{O}) \d F(\theta) \Big\} \nonumber\\
    &+\mathbb{E}\left[ \int \pardevt \left\{\prod_{t=1}^T \frac{I(A_t = d^{\theta}_t )}{\pi^t_{\epsilon}( A_t = d^{\theta}_t \mid \bar H_t )} \{Y-m(\theta,V)\}^2\right\}_{\epsilon=0} \d F(\theta) \right] \nonumber\\ 
      &= \mathbb{E}\Big\{\int \prodK\{Y - m(\theta,V)\}^2 \score(\mathcal{O}) \d F(\theta) \Big\} \nonumber\\
      &- \mathbb{E}\int \Big[ \sum_{t=1}^T \prod_{j\ne t} \frac{\Ij \It \{Y - m(\theta,V)\}^2}{\pi^t_0( A_j = d^{\theta}_j \mid \bar H_j ) \pi^t_0(A_t = d^{\theta}_t \mid \bar H_t)^2}\pardevt \pi_\epsilon(A_t = d^{\theta}_t|\bar H_t ) \d F(\theta) \Big] \label{eq:aug1}
\end{align}
We derive the form of (\ref{eq:aug1}) by focusing on a given $t$ which then easily generalizes to all $1 \leq t \leq T$. We have
 \begin{align}
    &\mathbb{E} \Big[\int \prod_{j \ne t}^T \Big \{ \frac{\Ij}{\pi^j_0( A_j = d^{\theta}_j |\bar H_j)} \Big\} \frac{\It}{\pi^t_0( A_t = d^{\theta}_t \mid \bar H_t)^2}\{Y - m(\theta,V)\}^2\pardevt \pi_\epsilon( A_t = d^{\theta}_t \mid \bar H_t) \d F(\theta) \Big] \nonumber\\
    &= \mathbb{E}\left[\int  \frac{ \It}{\pi^t_0 \left\{A_t = d^{\theta}_t \mid \bar H_t \right\}}\mathbb{E}\left[ \prod_{j \ne t}^T \frac{ \Ij  }{\pi^j_0\left\{ A_j = d^{\theta}_j  | \bar H_j \right\}}\{Y - m(\theta,V)\}^2  \mid  \bar S_t, \bar A_{t} \right] \score(\mathcal{O}) \d F(\theta) \right] \nonumber\\
    &- \mathbb{E}\Bigg[ \int \mathbb{E}\left[  \prod_{j \ne m}^T \frac{ \Ij  }{\pi^j_0\left\{ A_j = d^{\theta}_j  | \bar H_j \right\}}\{Y - m(\theta,V)\}^2  \mid  \bar S_t, \bar A_{t-1}, A_t = d^{\theta}_t \big] \score(\mathcal{O}) \d F(\theta) \right] \nonumber\\
    &= \mathbb{E}\left[ \int \frac{ \It}{\pi^t_0 \left\{A_t = d^{\theta}_t \mid \bar H_t \right\}}\mathbb{E}\left[ \prod_{j \ne m}^T \frac{ \Ij  }{\pi_0^j\left\{ A_j = d^{\theta}_j  | \bar H_j \right\}}\{Y - m(\theta,V)\}^2  \mid  \bar S_t, \bar A_{t-1}, A_t = d^{\theta}_t \right] \score(\mathcal{O}) \d F(\theta) \right] \nonumber\\
    &- \mathbb{E}\Bigg[ \int \mathbb{E}\left[  \prod_{j \ne m}^T \frac{ \Ij  }{\pi_0^j\left\{ A_j = d^{\theta}_j  | \bar H_j \right\}}\{Y - m(\theta,V)\}^2  \mid  \bar S_t, \bar A_{t-1}, A_t = d^{\theta}_t \big] \score(\mathcal{O}) \d F(\theta)\right] \nonumber\\
    &= \mathbb{E}\left[ \int \left\{\prod_{i = 1}^{t-1}\frac{I(A_i = d^{\theta}_i)}{\pi^t_0( A_i = d^{\theta}_i \mid \bar H_i)}\right\} Q^{t}_{m}\left[ \frac{\It}{\pi_0^t \left\{A_t = d^{\theta}_t \mid \bar H_t \right\}} - 1\right] \score(\mathcal{O}) \d F(\theta) \right] \nonumber\\
    &= \mathbb{E}\left[\int \left\{\prod_{i = 1}^{t-1}\frac{I(A_i = d^{\theta}_i)}{\pi^t_0( A_i = d^{\theta}_i \mid \bar H_i)}\right\} Q^t_{0,m}\left[\frac{\It- {\pi_0^t\left\{A_t = d^{\theta}_t \mid \bar H_t \right\}}}{{\pi_0^t\left\{A_t = d^{\theta}_t \mid \bar H_t \right\}}} \right] \score(\mathcal{O}) \d F(\theta) \right], \nonumber 
  \end{align}
  where $$Q^{t}_{0,m} = \mathbb{E}\left[\prod_{j = {t+1}}^T \frac{\Ij }{\pi_0^j\{ A_j = d^{\theta}_t \mid \bar H_j \}}\{Y - m(\theta,V)\}^2 \mid \bar H_t, A_t = d^{\theta}_t \right].$$
 Hence, the un-centered canonical gradient of $\Psi_{0, \text{nda}}(m)$ is 

 \begin{align}
     \varphi_m(\pi, Q_{0,m}) =& \int \left\{ 
            \prod_{t  = 1}^T \frac{ I(A_t = d_t^{\theta})}{\pi^t_0(A_t = d^{\theta}_t \mid \bar H_t ) }\{Y - m(\theta,V)\}^2 
            \right. \nonumber\\
            &\left. - \sum_{t = 1}^T \frac{I(A_t = d^{\theta}_t) - \pi^t_0(A_t = d^{\theta}_t \mid \bar H_t ) }{\pi^t_0 (A_t = d^{\theta}_t \mid \bar H_t)  }Q^t_{0,m} \prod_{j = 1}^{t-1} \frac{I(A_t = d_t^\theta)}{\pi^j_0 (A_t = d^{\theta}_t \mid \bar H_t ) } 
        \right\} \,dF(\theta).
 \end{align}

\section{Proof of Theorem 2}
\label{proof:thrm2}





We let $\pi_{n,b}$ be the estimate of $\pi_0$ obtained using the $b^{th}$ sample split. Similarly, $Q_{n,b,m}$ is the estimate of $Q_{0,m}$ and $m_{b}$ is the estimate of $m_0$ obtained using the $b^{th}$ sample split. 

\begin{align*}
    &\Psi^{MR}_{n}(m_n,\pi_n, Q_{n,m_n}) - \Psi_{0}(m_n) \\
   =&\Psi^{MR}_{n}(m_n,\pi_n, Q_{n,m_n}) - \Psi_{0}(m_n) \pm \Psi_{n}^{MR}(m_n,\pi_0, Q_{0,m_n})\nonumber \\ 
   =& \E \P^1_{n,b} \varphi_{m_{n,b}}(\pi_{n,b}, Q_{n,b, m_{n,b}}) - \E \P_0 \varphi_{m_{n,b}}(\pi_{0}, Q_{0,m_{n,b}}) 
   \pm \E \P^1_{n,b} \varphi_{m_{n,b}}(\pi_{0}, Q_{0,m_{n,b}}) \nonumber \\
   =& \E (\P^1_{n,b} - \P_0)\varphi_{m_{n,b}}\left(\pi_0, Q_{0, m_{n,b}}\right) + \E \P^1_{n,b} \varphi_{m_{n,b}}(\pi_{n,b}, Q_{n,b, m_{n,b}}) - \E \P^1_{n,b} \varphi_{m_{n,b}}(\pi_{0}, Q_{0,m_{n,b}}) \\
   =&\E (\P^1_{n,b} - \P_0)\varphi_{m_{n,b}}\left(\pi_0, Q_{0, m_{n,b}}\right) + \E \P^1_{n,b} \left\{ \varphi_{m_{n,b}}(\pi_{n,b}, Q_{n,b, m_{n,b}}) - \varphi_{m_{n,b}}(\pi_{0}, Q_{0,m_{n,b}}) \right\}\\
   =&\E (\P^1_{n,b} - \P_0)\varphi_{m^*}\left(\pi_0, Q_{0,m^*}\right) + \E \P_0 \left\{ \varphi_{m_{n,b}}(\pi_{n,b}, Q_{n,b, m_{n,b}}) - \varphi_{m_{n,b}}(\pi_{0}, Q_{0,m_{n,b}}) \right\} + o_p(n^{-1/2}) \\
   =&(\P_n - \P_0)\varphi_{m^*}\left(\pi_0, Q_{0,m^*}\right) + \E \P_0 \left\{ \varphi_{m_{n,b}}(\pi_{n,b}, Q_{n,b, m_{n,b}}) - \varphi_{m_{n,b}}(\pi_{0}, Q_{0,m_{n,b}}) \right\} + o_p(n^{-1/2})
\end{align*} 

We need to show that $\E \P_0 \left\{ \varphi_{m_{n,b}}(\pi_{n,b}, Q_{n,b, m_{n,b}}) - \varphi_{m_{n,b}}(\pi_{0}, Q_{0,m_{n,b}}) \right\} = o_p(n^{-1/2})$.

\begin{align*}
    &\E \P_0 \{\varphi_{m_{n,b}}(\pi_{n,b}, Q_{n,b,m_{n,b}}) - \varphi_{m_{n,b}}(\pi_0, Q_{0,m_{n,b}})\} \\
    =& \E \P_0 \int \left[ \prod_{t = 1}^T \frac{\it}{\pi^t_{n,b}}\{Y - m_{n,b}(\theta,V)\}^2 -\prod_{t=1}^T \frac{\it}{\pi^t_0}\{Y - m_{n,b}(\theta,V)\}^2\right]dF(\theta)\\
    &+ \sum_{t = 1}^T \E \P_0 \int \left[ \prod_{i = 1}^t \frac{\ii}{\pi^i_{0}}Q^t_{0, m_{n,b}} -  \prod_{i = 1}^t \frac{\ii}{\pi^i_{n,b}}Q^t_{n,b,m_{n,b}} \right]dF(\theta)\\
    &+ \sum_{t = 1}^T \E\P_0 \int \left[ \prod_{i = 1}^{t-1} \frac{\ii}{\pi^i_{n,b}}Q^{t}_{b,m_{n,b}} - \prod_{i = 1}^{t-1} \frac{\ii}{\pi^i_{0} }Q^{t}_{0,m_{n,b}}\right]dF(\theta) \\
    &= (I) + (II) + (III)
\end{align*}

For the rest of this calculation, we use the color brown to indicate that the term converges to 0 at the $n^{1/2}$ rate under the required assumptions.

\begin{align*}
    (I) =& \E \P_0 \int \prod_{t = 1}^T \frac{\it}{\pi^t_{n,b}}\{Y - m_{n,b}(\theta,V)\}^2 - \prod_{t = 1}^T \frac{\it}{\pi^t_{0}}\{Y - m_{n,b}(\theta,V)\}^2 dF(\theta) \\
    &\pm \sum_{t = 1}^T \E \P_0 \int \left[ \frac{\prod_{i=1}^T \ii}{\left( \prod_{i = 1}^{t-1} \pi^i_{n,b} \right)\left(\prod_{i = t}^T \pi^i_{0} \right)}\left\{Y - m_{n,b}(\theta, V)\right\}^2  \right]dF(\theta) \nonumber \\
    =& \sum_{i = 1}^T \E \P_0 \int \frac{\prod_{i=1}^T \ii}{\left( \prod_{i = 1}^t \pi^i_{n,b} \right)\left( \prod_{i = t+1}^T \pi^i_{0} \right)}\left\{Y - m_{n,b}(\theta, V)\right\}^2 \left\{\frac{\pi^t_0 - \pi^t_{n,b}}{\pi^t_0}\right\}dF(\theta)\\
    =& {\color{brown} \sum_{t=1}^T \sum_{k < l <m<t} \E \P_0  \int \frac{\prod_{i=1}^T \ii}{\pi^k_{n,b} \pi^m_{0}}\{Y - m_{n,b}(\theta,V)\}^2\frac{\pi_{l,0} - \pi_{l,S_n}}{\pi^t_0\pi^t_{n,b}}\frac{\pi^t_0 - \pi^t_{n,b}}{\pi^t_0\pi^t_{n,b}}dF(\theta) }\\
    &+ \sum_{t=1}^T \E \P_0 \int \left(\prod_{i=1}^T \frac{\ii}{\pi^i_{0}}\left\{Y - m_{n,b}(\theta,V) \right\}^2 \frac{\pi^t_0 - \pi^t_{n,b}}{\pi^t_{n,b}}\right)dF(\theta) \\
    =&  \sum_{t=1}^T \E \P_0 \int \left(\prod_{i=1}^T \frac{\ii}{\pi^i_{0}}\left\{Y - m_{n,b}(\theta, V)\right\}^2 \frac{\pi^t_0 - \pi^t_{n,b}}{\pi^t_{n,b}}\right)dF(\theta) + o_p(n^{-1/2}) \\
    =& \sum_{t=1}^T \E \P_0 \int \prod_{i=1}^t \frac{\ii}{\pi^i_{0}}\left(Q^t_{0,m_{n,b}}\right)\frac{\pi^t_0 -  \pi^t_{n,b}}{\pi^t_{n,b}} dF(\theta) + o_p(n^{-1/2})
\end{align*}

We continue with Term (II)

\begin{align*}
    (II) =& \sum_{t = 1}^T \E \P_0 \int \left[ \prod_{i = 1}^t \frac{\ii}{\pi^i_{0}}Q^t_{0,m_{n,b}} -  \prod_{i = 1}^t \frac{\ii}{\pi^i_{n,b}}Q^t_{n,b,m_{n,b}} \right]dF(\theta)\\
    =& \sum_{t = 1}^T \E \P_0 \int \left[ \prod_{i=1}^t \frac{\ii}{\pi^i_{0}}Q^t_{0,m_{n,b}} -  \prod_{i = 1}^t \frac{\ii}{\pi^i_{n,b}}Q^t_{n,b,m_{n,b}} \right]dF(\theta) \\
    &\pm \sum_{t=1}^T \sum_{k = 1}^t \E \P_0 \int  \frac{\prod_{i=1}^t \ii}{\prod_{i = 1}^{k-1} \pi^i_{n,b} \prod_{i=k}^t \pi^i_{0} }Q^t_{n,b,m_{n,b}} \\
    =& \sum_{t = 1}^T \E \P_0 \int \prod_{i=1}^t\frac{\ii}{\pi^i_{0}}\left( Q^t_{0,m_{n,b}} - Q^t_{n,b,m_{n,b}}\right)dF(\theta) \\
    &+ {\color{brown} \sum_{t = 1}^T \sum_{i=1}^t \E \P_0 \int \frac{\prod_{i=1}^t \ii}{ \prod_{k=1}^{i-1} \left( \pi^k_{n,b} \right) \prod_{l= i }^t \left( \pi^l_{0} \right)}\left(Q^t_{n,b,m_{n,b}} -Q^t_{0,m_{n,b}} \right)\left( \frac{\pi^i_{n,b} - \pi^i_0}{\pi^i_{n,b}} \right)dF(\theta) } \\
    &+ \sum_{t = 1}^T \sum_{i = 1}^t \E \P_0 \int \frac{\prod_{i = 1}^t \ii}{\prod_{k = 1}^{i-1} \left(\pi^k_{n,b} \right) \prod_{l=i}^t \left( \pi^l_{0} \right)}\left( \frac{\pi^i_{n,b} - \pi^i_{0}}{\pi^i_{n,b}}\right)Q^t_{0,m_{n,b}} dF(\theta)\\
    =& \sum_{t = 1}^T \E \P_0 \int \prod_{i=1}^t\frac{\ii}{\pi^i_{0}}\left( Q^t_{n,b,m_{n,b}} - Q^t_{0,m_{n,b}}\right)dF(\theta) \\
    &+ \sum_{t = 1}^T \sum_{i = 1}^t \E \P_0 \int \frac{\prod_{i = 1}^t \ii}{\prod_{i =1}^t \left(\pi^i_{0} \right)}\left( \frac{\pi^i_{n,b} - \pi^i_{0}}{\pi^i_{n,b}}\right)Q^t_{0,m_{n,b}} dF(\theta) \\
    &+ {\color{brown} \sum_{t=1}^T \sum_{1 \le k \le l \le t} \E \P_0 \int \frac{\prod_{i =1}^t \ii}{ \prod_{i=1}^t \left( \pi^i_{0} \right)}\left(\frac{\pi^l_{n,b} - \pi^l_{0}}{\pi^l_{n,b} } \right)\left(\frac{\pi^k_{0} - \pi^k_{n,b}}{\pi^k_{n,b} } \right)Q^t_{0,m_{b}} dF(\theta)} + o_p(n^{-1/2})  \\
     =& \sum_{t = 1}^T \E \P_0 \int \prod_{i=1}^t\frac{\ii}{\pi^i_{0}}\left( Q^t_{n,b,m_{b}} - Q^t_{0,m_{b}}\right)dF(\theta) \\
    &+ \sum_{t = 1}^T \sum_{i = 1}^{t} \E \P_0 \int \frac{\prod_{i = 1}^t \ii}{\prod_{i =1}^t \left(\pi^i_{0} \right)}\left( \frac{\pi^i_{n,b} - \pi^i_{0}}{\pi^i_{n,b}}\right)Q^t_{0,m_{b}} dF(\theta) + o_p(n^{-1/2})
\end{align*}

Lastly, we consider the third term in red. 

\begin{align*}
    (III) =& \sum_{t=1}^T \E \P_0 \int \left[\prod_{i =1}^{t-1} \frac{\ii}{\pi^i_{n,b}}Q^t_{n,b,m_{n,b}} - \prod_{i=1}^{t-1} \frac{\ii}{\pi^i_{n,b}}Q^t_{0,m_{n,b}} \right]dF(\theta) \\
     =& \sum_{t=1}^T \E \P_0 \int \left[\prod_{i =1}^{t-1} \frac{\ii}{\pi^i_{n,b}}Q^t_{n,b,m_{n,b}} - \prod_{i=1}^{t-1} \frac{\ii}{\pi^i_{n,b}}Q^t_{0,m_{n,b}} \right]dF(\theta) \\
    &\pm \sum_{t=1}^T \sum_{k = 1}^{t-1} \E \P_0 \int  \frac{\prod_{i=1}^{t-1} \ii}{\prod_{i = 1}^{k-1} \pi^i_{n,b} \prod_{i=k}^{t-1} \pi^i_{0} }Q^t_{n,b,m_{n,b}} \\
    =& \sum_{t = 1}^T \E \P_0 \int \prod_{i=1}^{t-1}\frac{\ii}{\pi^i_{0}}\left( Q^t_{n,b,m_{n,b}} -Q^t_{0,m_{n,b}} \right)dF(\theta) \\
    &+ {\color{brown} \sum_{t = 1}^T \sum_{i =1}^{t-1} \E \P_0 \int \frac{\prod_{i=1}^{t-1} \ii}{ \prod_{k=1}^i \left( \pi^k_{n,b} \right) \prod_{l=i}^{t-1} \left( \pi^i_{0} \right)}\left(Q^t_{0,m_{n,b}} -Q^t_{n,b,m_{n,b}} \right)\left( \frac{\pi^l_{0} - \pi^l_{n,b}}{\pi^l_{n,b}} \right)dF(\theta) } \\
    &+ \sum_{t = 1}^T \sum_{1 \le l \le t-1} \E \P_0 \int \frac{\prod_{i = 1}^{t-1} \ii}{\prod_{k=1}^i \left(\pi^i_{n,b} \right) \prod_{l=i}^{t-1} \left( \pi^i_0 \right)}\left( \frac{\pi^l_0 - \pi^l_{n,b}}{\pi^l_{n,b}}\right)Q^t_{0,m_{n,b}} dF(\theta)\\
    =& \sum_{t = 1}^T \E \P_0 \int \prod_{i=1}^{t-1}\frac{\ii}{\pi^i_{0}}\left( Q^t_{0,m_{n,b}} - Q^t_{n,b,m_{n,b}}\right)dF(\theta) \\
    &+ \sum_{t = 1}^T \sum_{i = 1}^{t-1} \E \P_0 \int \frac{\prod_{i = 1}^t \ii}{\prod_{i =1}^t \left(\pi^i_{0} \right)}\left( \frac{\pi^i_0 - \pi^i_{n,b}}{\pi^i_{n,b}}\right)Q^t_{0,m_{n,b}} dF(\theta) \\
    &+ { \color{brown} \sum_{t=1}^T \sum_{1 \le k \le l \le t-1} \E \P_0 \int \frac{\prod_{i =1}^{t-1} \ii}{ \prod_{i=1}^{t-1} \left( \pi^i_{0} \right)}\left(\frac{\pi^l_0 - \pi^l_{n,b}}{\pi^l_{n,b} } \right)\left(\frac{ \pi^k_{n,b} - \pi^k_{0}}{\pi^k_{n,b} } \right)Q^t_{n,b,m_{n,b}} dF(\theta)} + o_p(n^{-1/2}) \\
    =& \sum_{t = 1}^T \E \P_0 \int \prod_{i=1}^{t-1}\frac{\ii}{\pi^i_{0}}\left( Q^t_{0,m_{n,b}} - Q^t_{n,b,m_{n,b}}\right)dF(\theta) \\
    &+ \sum_{t = 1}^T \sum_{i = 1}^{t-1} \E \P_0 \int \frac{\prod_{i = 1}^{t-1} \ii}{\prod_{i =1}^{t-1}  \pi^i_{0} }\left( \frac{\pi^i_{0} - \pi^i_{n,b}}{\pi^i_{n,b}}\right)Q^t_{0,m_{n,b}} dF(\theta) + o_p(n^{-1/2}) 
\end{align*}

We combine all the remaining terms in (I), (II), and (III)

\begin{align*}
    &(I) + (II) + (III)  \\
    &= \sum_{t = 1}^T \E \P_0 \int \prod_{i=1}^t\frac{\ii}{\pi^i_{0}}\left( Q^t_{n,b,m_{n,b}} - Q^t_{0,m_{n,b}}\right)\left(\frac{\it - \pi^t_0}{\pi^t_0} \right)dF(\theta) \\
    &+ \sum_{t = 1}^T \sum_{i = 1}^{t} \E \P_0 \int \frac{\prod_{i = 1}^t \ii}{\prod_{i =1}^t \left(\pi^i_{0} \right)}\left( \frac{\pi^i_{n,b} - \pi^i_{0}}{\pi^i_{n,b}}\right)Q^t_{0,m_{n,b}} dF(\theta) + o_p(n^{-1/2}) \\
    &+  \sum_{t = 1}^T \sum_{i = 1}^{t} \E \P_0 \int \frac{\prod_{i = 1}^t \ii}{\prod_{i =1}^t \left(\pi^i_{0} \right)}\left( \frac{\pi^i_{0} - \pi^i_{n,b}}{\pi^i_{n,b}}\right)Q^t_{0,m_{n,b}} dF(\theta) + o_p(n^{-1/2})\\
    &= o_p(n^{-1/2})
\end{align*}

Hence, 

\begin{equation*}
     \Psi^{MR}_n(m_n,\pi_n, Q_{n,m_n}) - \Psi_{0}(m_n) = (\P_n - \P_0)\varphi_{m^*}(\pi_0, Q_{0,m^*}) + o_p(n^{-1/2})
\end{equation*}

\section{Proof of Theorem 3}
\label{proof:thrm3}

In the calculation below, we show that if we look at $\Psi^{MR}_{n}(m_n,\pi_n, Q_{m_n,n}) -  \Psi_{0}(m^*)$ instead of $\Psi^{MR}_{n}(m_n,\pi_n, Q_{m_n,n}) - \Psi_{0}(m_n)$, we will have a bias term. 

\begin{align*}
    \Psi^{MR}_{n}&(m_n, \pi_n, Q_{n,m_n}) - \Psi_0(m^*) \\
    =&  \Psi^{MR}_{n}(m_n, \pi_n, Q_{n, m_n}) - \Psi_0(m^*) \pm \Psi^{MR}_n(m^*, \pi_0, Q_{0,m^*})\\
    =& \E \P_{B_n} \varphi_{m_{n,b}}(\pi_{n,b}, Q_{n,b,m_{n,b}}) - \P_0 \varphi_m^*(\pi_0, Q_{0,m^*}) \pm \E \P_n \varphi_m^*(\pi_0, Q_{0,m^*} )\\
    =& (\P_n - \P_0)\varphi_{m^*}(\pi_0, Q_{0,m^*})+ \E \P_{B_n} \left\{ \varphi_{m_{n,b}}( \pi_{n,b}, Q_{n,b,m_{n,b}}) - \varphi_{m^*}(\pi_0, Q_{0,m^*}) \right\}\\
    &\pm \E \P_{B_n} \varphi_{m_{n,b}} (\pi_0, Q_{0, m_{n,b}}) \\ 
    =& (\P_n - \P_0)\varphi_{m^*}(\pi_0, Q_{0,m^*}) + \E \P_{B_n}\left\{ \varphi_{m_{n,b}} (\pi_{n,b}, Q_{n,b,m_{n,b}}) - \varphi_{m_{n,b}}(\pi_0, Q_{0, m_{n,b}}) \right\}\\
    &+ \E \P_{B_n} \left\{\varphi_{m_{n,b}}(\pi_0, Q_{0, m_{n,b}}) - \varphi_{m^*}(\pi_0, Q_{0,m^*}) \right\} \\
    =& (\P_n - \P_0)\varphi_{m^*}(\pi_0, Q_{0,m^*}) + \E \P_0 \left\{\varphi_{m_{n,b}} (\pi_{n,b}, Q_{n,b,m_{n,b}}) - \varphi_{m_{n,b}}( \pi_0, Q_{0,m_{n,b}})\right\} \\
    &+ \E \P_0\{\varphi_{m_{n,b}}(\pi_0, Q_{0, m_{n,b}}) - \varphi_{m^*}(\pi_0, Q_{0,m^*})\} 
\end{align*}
\\
In the proof of Theorem 2, we have shown that \\ $\E \P_0 \left\{\varphi_{m_{n,b}} (\pi_{n,b}, Q_{n,b,m_{n,b}}) - \varphi_{m_{n,b}}( \pi_0, Q_{0,m_{n,b}})\right\} = o_p(n^{-1/2})$. We will continue to examine the third term. 

{\small
\begin{align*}
    \E \P_0 &[ \varphi_{m_{n,b}}(\pi_0, Q_{0, m_{n,b}}) - \varphi_{m^*}(\pi_0, Q_{0,m^*}) ]\\
    =& \E \int \P_0 \prod_{t=1}^T \frac{\it}{\pi^t_0}\left\{ \{Y - m_{n,b}(\theta,V)\}^2 - (Y - m^*(\theta, V))^2 \right\}dF(\theta) \\
    =& \E \int \P_0 \prod_{t=1}^T \frac{\it}{\pi^t_0}\left\{ Y^2 - 2Ym_{n,b}(\theta, V) + m_{n,b}^2(\theta, V) - Y^2 + 2Ym^*(\theta, V) - {m^*(\theta, V)}^2 \right\}dF(\theta)\\
     =& \E \int \P_0 \prod_{t=1}^T \frac{\it}{\pi^t_0}\{ Y^2 - 2m_0(\theta,V)m_{n,b}(\theta, V) + m_{n,b}^2(\theta, V) - Y^2 \\
     &\hspace{4.5cm}+ 2m_0(\theta,V)m^*(\theta, V) - {m^*(\theta, V)}^2 \}dF(\theta)\\
     =& \E \int \P_0 \prod_{t=1}^T \frac{\it}{\pi^t_0}\left[\{m_0(\theta,V) - m_{n,b}\}^2 - \{m^*(\theta, V) - m_0(\theta,V)\}^2 \right]dF(\theta) \\
     =& \E \int \P_0 \prod_{t=1}^T \frac{\it}{\pi^t_0}\big[ \{m_0(\theta,V) - m^*(\theta, V) + m^*(\theta, V) - m_{n,b}(\theta, V)\}^2 \\
     &\hspace{4.5cm}- \{m^*(\theta, V) - m_0(\theta,V)\}^2 \big]dF(\theta)\\
     =& \E \int \P_0 \prod_{t=1}^T \frac{\it}{\pi^t_0}\big[ \{m^*(\theta, V) - m_{n,b}(\theta, V)\}^2 \\
     &\hspace{4.5cm}- 2\{m^*(\theta, V) - m_0(\theta,V)\}\{m^*(\theta, V) - m_{n,b}(\theta, V)\} \big]dF(\theta)
\end{align*}
}

where $m_0(\theta,V) = E[Y|\theta, V]$ is the true minimizer of the risk function and  $m_n(\theta, V) \xrightarrow{} m^*(\theta, V)$. 

Hence $\E \P_0 [ \varphi_{m_{n,b}}(\pi_0, Q_{0, m_{n,b}}) - \varphi_{m^*}(\pi_0, Q_{0,m^*}) ] = o_p(n^{-1/2})$ if and only if $m_0(\theta,V) = m^*(\theta, V)$.

\section{Proof of Theorem 4}
\label{proof:thrm4}

In this section, we will show that by undersmoothing the propensity score, the IPW estimator will be linear asymptotic. We start by considering the difference below.  
\begin{align}
\Psi^{UIPW}_n(m_n,\pi_n)-\Psi_0(m_n) =&\E \P_0 \left\{L_{m_{n,b}}\left(\pi_{n, b,\lambda} \right)-L_{m_{n,b}}\left(\pi_0 \right)\right\}   \nonumber \\
&+ \E (\P^1_{n,b} - \P_0)  L_{m_{n,b}}\left(\pi_{n, b,\lambda} \right) \label{eq:1}
\end{align}

The second term on the right-hand side of (\ref{eq:1}) can be represented as
\begin{align}
\E (\P^1_{n,b} - \P_0) L_{m_{n,b}}\left(\pi_{n, b,\lambda} \right) =&\E (\P^1_{n,b} - \P_0) L_{m_{n,b}}\left(\pi_{n, b,\lambda} \right)-L_{m^*}\left(\pi_0 \right) \nonumber \\
&+\E (\P_{n, b}^1-\P_0)  L_{m^*}\left(\pi_0 \right) \\  \label{eq:2}
=& \E (\P_{n}-\P_0)  L_{m^*}\left(\pi_0 \right) + o_p(n^{-1/2}) 
\end{align}

The only remaining term to study is the first term in (\ref{eq:1}).

$$
\begin{aligned}
\E &\P_0  \left\{L_{m_{n,b}}\left(\pi_{n, b,\lambda} \right)-L_{m_{n,b}}\left(\pi_0 \right)\right\}\\
= & \E\P_0 \left\{L_{m_{n,b}}\left(\pi_{n, b,\lambda} \right)-L_{m_{n,b}}\left(\pi_0 \right)\right\} 
 -\E \P_0\left\{L_{m^*}\left(\pi_{n, b,\lambda} \right)-L_{m^*}\left(\pi_0 \right)\right\}  \\
 &+\E \P_0\left\{L_{m^*}\left(\pi_{n, b,\lambda} \right)-L_{m^*}\left(\pi_0 \right)\right\} 
\end{aligned}
$$
The terms in the first integral can be written as

\begin{align*}
&\E \P_0 \{  L_{m_{n,b}}\left(\pi_{n, b,\lambda} \right)-L_{m_{n,b}}\left(\pi_0 \right) \} =\E \int \int \mathbb{E}\left\{\left(Y-m_{n,b}(\theta, V)\right)^2 \mid \bar H_T\right\} \pi_0\left(\frac{\pi_0-\pi_{n, b,\lambda}}{\pi_{n, b,\lambda} \pi_0}\right) dF(\theta) dP\left(o\right) \\
&\E \P_0 \{ L_{m^*}\left(\pi_{n, b,\lambda} \right) - L_{m^*}\left(\pi_0 \right) \} = \E \int \int \mathbb{E}\left\{\left(Y-m^*(\theta,V)\right)^2 \mid \bar H_T\right\}\pi_0\left(\frac{\pi_0-\pi_{n, b,\lambda}}{\pi_{n, b,\lambda} \pi_0}\right) dF(\theta) dP\left(o\right) .
\end{align*}

where $\pi_0 = \prod_{t=1}^T \pi_0^t$ and $\pi_{n,b}$ is an estimate of $\pi_0$ using the $b^{th}$ sample split.

Combining these two terms, we have
\begin{align*}
    &\E \left\{L_{m_{n,b}}\left(\pi_{n, b,\lambda}\right)-L_{m_{n,b}}\left(\pi_0 \right)\right\}-\left\{L_{m^*}\left(\pi_{n, b,\lambda} \right)-L_{m^*}\left(\pi_0 \right)\right\} = \\
&\int \int \left[ \{m^{*}(\theta,V) - m_{n, b}(\theta,V)\}\{2Y - m_{n,b}(\theta,V) - m^*(\theta,V)\} \right] \pi_0\left(\frac{\pi_0-\pi_{n, b,\lambda}}{\pi_{n, b,\lambda} \pi_0}\right) dF(\theta) d P\left(o \right) 
=o_p\left(n^{-1 / 2}\right).
\end{align*}
The last equality applies Cauchy-Schwarz inequality and the assumption that $\int \|\pi_n-\pi_0\|_2. \|m_n-m^*\|_2 dF(\theta) =o_p(n^{-1/2})$. Gathering the relevant terms, we have

\begin{align*}
\Psi^{UIPW}_n(m_n,\pi_n)-\Psi_0(m_n)=&\frac{1}{n} \sum_{i=1}^n\left\{L_{m^*}\left(\pi_0\right)-\Psi_0(m^*)\right\}\\
&+\E \P_0 \int\left\{L_{m^*}\left(\pi_{n, b,\lambda} \right)-L_{m^*}\left(\pi_0 \right)\right\} +o_p\left(n^{-1/2}\right).
\end{align*}

We now consider the second term on the right-hand side.
\begin{align}
    &\E \P_0 \left\{L_{m^*}\left(\pi_{n, b,\lambda} \right)-L_{m^*}\left(\pi_0 \right)\right\}  \nonumber \\
    &= \E \P_0 \int \prod_{t=1}^T \frac{I(A_t = d^{\theta}_t)\{Y - m^*(\theta,V)\}^2}{\pi^t_{n,b, \lambda}}dF(\theta) - \P_0 \int \prod_{t=1}^T \frac{I(A_t = d^{\theta}_t)\{Y - m^*(\theta,V)\}^2}{\pi^t_{0}}dF(\theta) \nonumber\\
    &\pm \sum_{k=2}^{T} \E\P_0 \int \frac{\prod_{t=1}^T I(A_t = d^{\theta}_t) \{Y - m^*(\theta,V)\}^2 }{\prod_{i=1}^{k-1}(\pi^t_{n,b, \lambda}) \prod_{t = k}^T(\pi^t_0)}  dF(\theta) \nonumber\\
    &=\sum_{k=1}^T \E \P_0 \int \frac{\prod_{t=1}^T I(A_t = d_t^{\theta})\{Y - m^*(\theta,V) \}^2}{\prod_{t=1}^{k}(\pi^t_{n,b, \lambda}) \prod_{t = k+1}^T (\pi^t_0) } dF(\theta) \nonumber\\
    &= \sum_{k=1}^T \E\P_0 \int \frac{\prod_{t=1}^{k-1} I(A_t = d^\theta_t)Q^k_{0, m^*} }{\prod_{t=1}^{k-1} (\pi_{t,b}) }\frac{\pi^k_{0}(\pi^k_{0} - \pi^k_{n,b, \lambda})}{\pi^k_{0}\pi^k_{n,b, \lambda}}dF(\theta) \nonumber\\
    &= \sum_{k=1}^T \E \P_0 \int \frac{\prod_{t=1}^{k -1}I(A_t = d^\theta_t)Q^k_{0, m^*}}{\prod_{t=1}^{k -1}(\pi^t_{n,b, \lambda})}\frac{\pi^k_{0} - \pi^k_{n,b, \lambda} }{\pi^k_{0}} - \frac{\prod_{t=1}^{k -1}I(A_t = d^\theta_t)Q^k_{0, m^*}}{\prod_{t=1}^{k -1}(\pi^t_{n,b, \lambda})}\frac{ (\pi^k_{0} - \pi^k_{n,b, \lambda})^2 }{\pi^k_{n,b, \lambda}} dF(\theta) \nonumber\\
    &= \sum_{k=1}^T \E \P_0 \int \frac{\prod_{t=1}^{k -1}I(A_t = d^\theta_t)Q^k_{0, m^*}}{\prod_{t=1}^{k -1}(\pi^t_{n,b, \lambda})}\frac{\pi^k_{0} - \pi^k_{n,b, \lambda} }{\pi^k_{0}} + o_p(n^{-1/2}) \nonumber\\
    &= \sum_{k=1}^T \E \P_0 \int \frac{\prod_{t=1}^{k -1}I(A_t = d^\theta_t)Q^k_{0, m^*}}{\prod_{t=1}^{k -1}(\pi^t_{n,b, \lambda})}\frac{ I(A_k =d^{\theta}_k) - \pi^k_{n,b, \lambda} }{\pi^k_{0}} + o_p(n^{-1/2}). \label{eq:dcar1} 
\end{align}

Let 
\begin{align}
    &D_{CAR}^k(Q_0, \pi_0, \pi_{n, b,\lambda}) 
    = \int \frac{\prod_{t=1}^{k -1}I(A_t = d^\theta_t)Q^k_{0,m^*}}{\prod_{t=1}^{k -1}\pi^t_{n,b, \lambda}}\frac{I(A_k =d^{\theta}_k) - \pi^k_{n,b, \lambda} }{\pi^k_{0}} \\
    &D_{CAR}^k(Q_0, \pi_0) = \int \frac{\prod_{t=1}^{k -1}I(A_t = d^\theta_t)Q^k_{0, m^*}}{\prod_{t=1}^{k -1}\pi^t_{n,b, \lambda}}\frac{I(A_k =d^{\theta}_k) - \pi^k_{0} }{\pi^k_{0}}
\end{align}
and $D_{CAR}(Q_0, \pi_0, \pi_{n, b,\lambda}) = \sum_{k = 1}^T D_{CAR}^k(Q_0, \pi_0, \pi_{n, b,\lambda})$ and $D_{CAR}(Q_0, \pi_0) = \sum_{k = 1}^T D_{CAR}^k(Q_0, \pi_0)$. We then have
\begin{align*}
    (\ref{eq:dcar1})&= \E \P_0 D_{CAR}(Q_0, \pi_0, \pi_{n, b,\lambda}) + o_p(n^{-1/2}) \\
    =& \E \P_0 D_{CAR}(Q_0, \pi_0, \pi_{n, b,\lambda}) - \E \P_{n, b}^1 D_{CAR}(Q_0, \pi_0, \pi_{n, b,\lambda}) + \E \P_{n, b}^1 D_{CAR}(Q_0, \pi_0, \pi_{n, b,\lambda}) + o_p(n^{-1/2}) \\
    =& \E (\P_0 - \P_{n, b}^1) D_{CAR}(Q_0, \pi_0, \pi_{n, b,\lambda}) + \E \P_{n, b}^1 D_{CAR}(Q_0, \pi_0, \pi_{n, b,\lambda}) + o_p(n^{-1/2}) \\
    =& \E (\P_0 - \P_{n, b}^1) [D_{CAR}(Q_0, \pi_0, \pi_{n, b,\lambda}) - D_{CAR}(Q_0, \pi_0)] +   \E (\P_0 - \P_{n, b}^1)D_{CAR}(Q_0, \pi_0) \\
    &+ \E \P_{n, b}^1 D_{CAR}(Q_0, \pi_0, \pi_{n, b,\lambda})+ o_p(n^{-1/2})\\
    =& (\P_0 - \P_n)D_{CAR}(Q_0, \pi_0)+ \E \P_{n, b}^1 D_{CAR}(Q_0, \pi_0, \pi_{n, b,\lambda})+ o_p(n^{-1/2}).
\end{align*}


Putting everything together, we have 
\begin{align}
    \Psi^{UIPW}_n(m_n,\pi_n)-\Psi_0(m_n,\pi_0) =& (\P_n-\P_0)\{L_m^*(\pi_0)-D_{CAR}(Q_0, \pi_0)\} \nonumber\\
    & + \sum_{k=1}^T \E \P_{n, b}^1 D^k_{CAR}(Q_0, \pi_0, \pi_{b,\lambda_{opt}})+o_p(n^{-1/2}),\\
    =& (\P_n-\P_0)\varphi_{m^*}(\pi_0, Q_{0,m^*})
\end{align}
so it remains to show the second term on the right-hand side is asymptotically negligible. 

For any given $k$ and any given collection of folds, we have 
\begin{align*}
    \P_{n, b}^1 D^k_{CAR}(Q_0, \pi_0, \pi_{b,\lambda_{opt}})
    & = \P_{n, b}^1 \int \frac{\prod_{t=1}^{k -1}I(A_t = d^\theta_t)Q^k_{0, m^*}}{\prod_{t=1}^{k -1}(\pi^t_{b, \lambda_{opt}})}\frac{I(A_k =d^{\theta}_k) - \pi^k_{b, \lambda_{opt}} }{\pi^k_{0}} \\
    & = o_p(n^{-1/2}),
\end{align*}
by applying Lemma 1 in \citet{ertefaie2023nonparametric}, which relies on our assumptions \dots [his asp 1 and 2 plus his equation 4, at each k] \dots. The conclusion that 
\begin{equation*}
     \Psi^{UIPW}_n(m_n, \pi_{n, \lambda_{opt}})-\Psi_0(m_n) = (\P_n-\P_0)\varphi_{m^*}(\pi_0, Q_{0,m^*}) + o_p(n^{-1/2}) 
\end{equation*}
followed by summing over $k$. 

\section{Additional Simulations} 
\label{sec:addsim}

\begin{figure}[ht]
    \centering
    \includegraphics[width=1\linewidth]{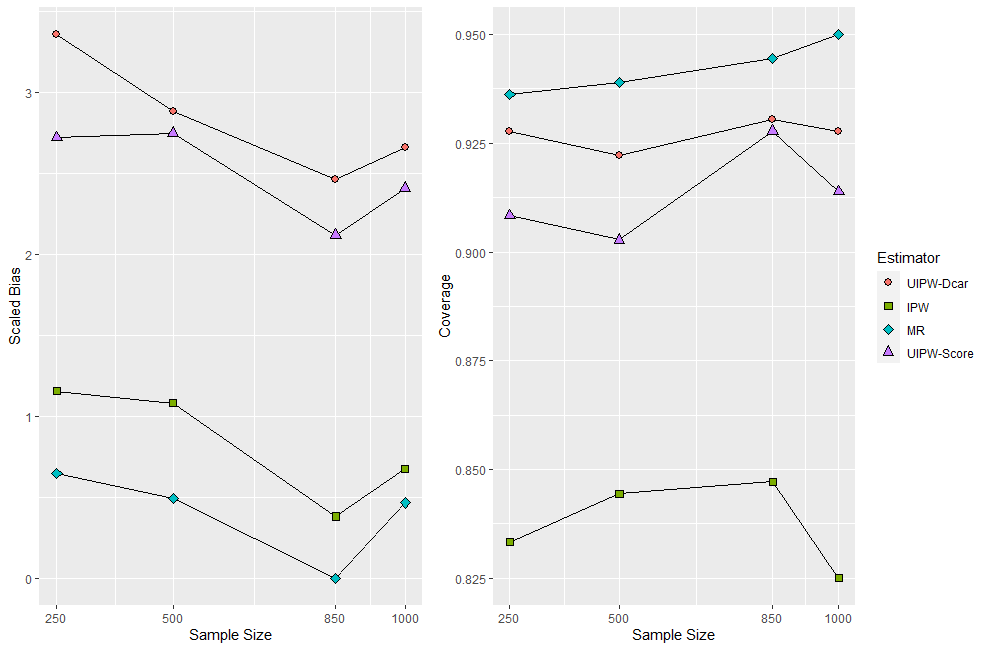}
    \caption{Randomized Treatment: Scale Bias and Coverage Plots}
    \label{fig:randomize-scenario}
\end{figure}

 We consider the case where the treatment $A_t$ is randomized (Figure \ref{fig:randomize-scenario}). Figure \ref{fig:randomize-scenario} shows that our proposed estimators and IPW estimator are all asymptotically linear. However, the IPW estimator is not efficient. Our proposed estimator is efficient. Their coverage all get close to 95\% as the sample size increases. The coverage of IPW estimator, on the other hand, decreases to 80\% as the sample size increases.  

We also present the simulation scenarios where the marginal structural models are estimated using

\begin{figure}[ht]
    \centering
    \includegraphics[width=1\linewidth]{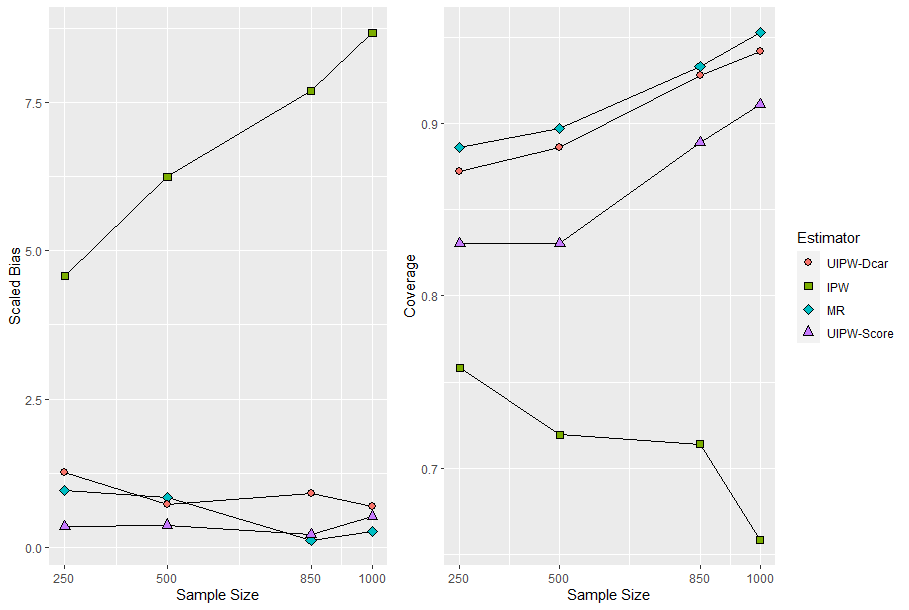}
    \caption{The marginal structural model is estimated using the linear regression with a second model term}
    \label{fig:m-lm-sq}
\end{figure}

\begin{figure}[ht]
    \centering
    \includegraphics[width=1\linewidth]{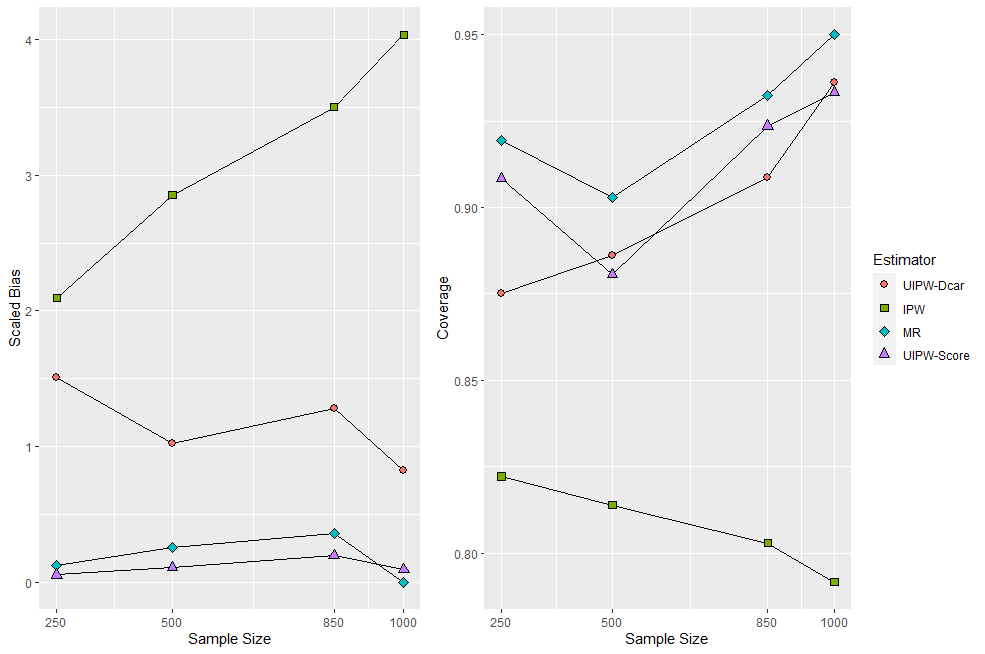}
    \caption{The marginal structural model is estimated using the random forest}
    \label{fig:m-rf}
\end{figure}

\begin{figure}[ht]
    \centering
    \includegraphics[width=1\linewidth]{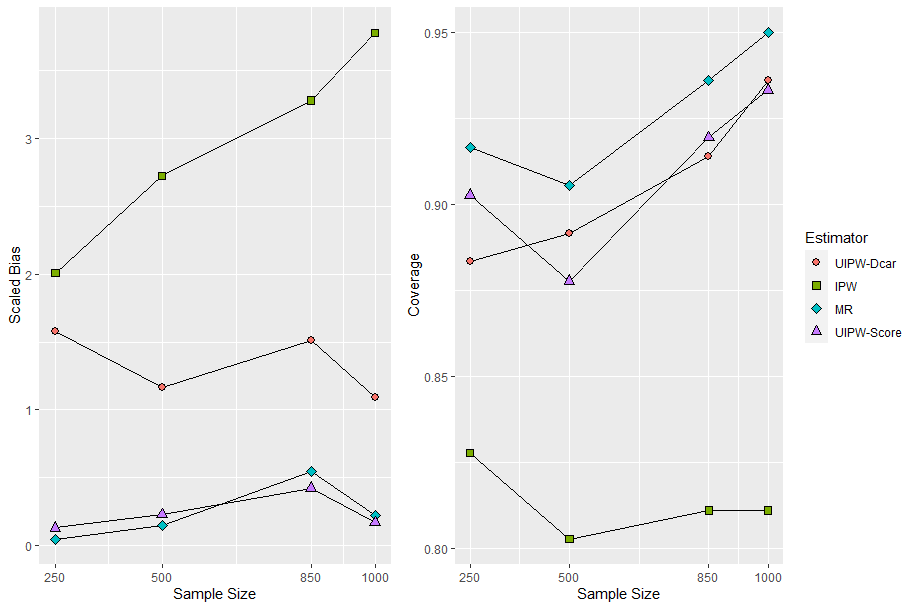}
    \caption{The marginal structural model is estimated using the highly adaptive lasso model}
    \label{fig:m-hal}
\end{figure}

\begin{enumerate} 
    \item Linear model with second order term (i.e, $E[Y\mid \theta] = \beta_0 + \beta_1\theta + \beta_2\theta^2$) (Figure \ref{fig:m-lm-sq})
    \item Random forest (Figure \ref{fig:m-rf})
    \item Highly adaptive lasso (Figure \ref{fig:m-hal})
\end{enumerate}

While our proposed estimators may display some variability in scaled bias in small sample sizes when the marginal structural model is estimated using nonparametric methods, they exhibit a downward trend as sample sizes increase, suggesting the asymmetrical linearity property. In contrast, the inverse probability weighting (IPW) estimator shows an upward trend, indicating potential bias when the propensity model is estimated using nonparametric methods. Notably, the UIPW-score estimator outperforms the UIPW-Dcar in various scenarios. Its scaled bias is consistently smaller than that of UIPW-Dcar and, in some instances, even smaller than the MR estimator. Furthermore, its coverage is closer to 0.95 compared to UIPW-Dcar in smaller sample sizes, especially when the marginal structural model is estimated using nonparametric methods.  
  
\end{document}